\documentclass[10pt,twocolumn,showpacs,eqsecnum,floatfix]{revtex4}

\usepackage{amsmath, amssymb, revsymb}
\usepackage{mathptmx}
\usepackage{enumerate}
\usepackage{graphicx}
\usepackage{subfigure}

\newcommand*{\mean}[1]{\langle #1 \rangle}
\newcommand*{\meanat}[2]{\mean{#1}_{#2}}
\newcommand*{\msd}{\mean{\Delta x^2}}  % mean square displacement
\newcommand*{\defeq}{:=}
\newcommand*{\eqdef}{=:}
\newcommand*{\defn}[1]{\textsl{#1}}
\newcommand*{\rd}{\, \textrm{d}}

\newcommand*{\pd}[2]{\frac{\partial #1}{\partial #2}}

\newcommand*{\from}{\colon}
\newcommand*{\M}{\mathcal{M}}

\newcommand*{\figref}[1]{Fig.~\ref{#1}}
\newcommand*{\bfigref}[1]{Figure~\ref{#1}}  % use at beginning of sentence
\newcommand*{\secref}[1]{Sec.~\ref{#1}}
\newcommand*{\bsecref}[1]{Section~\ref{#1}}

\newcommand*{\emp}{^{\mathrm{emp}}}

\newcommand*{\flow}{\Phi}

\newcommand*{\prob}[1]{\P \left( #1 \right)}

\newcommand*{\conddensity}[3]{#1\left(#2 \, \big| \, #3\right)}

\newcommand*{\D}{\Delta}
\newcommand*{\Dx}{\D x}
\newcommand*{\DX}{\D X}

\newcommand*{\rad}{^{\text{rad}}}
\newcommand*{\phirad}{\phi\rad}
\newcommand*{\rhorad}{\tilde{\rho}\rad}
\newcommand*{\ftrad}{\tilde{f}\rad}
\newcommand*{\frad}{f\rad}

\newcommand*{\distconv}{\stackrel{\mathcal{D}}{\longrightarrow}}

\newcommand*{\e}{\mathrm{e}}
\renewcommand*{\i}{\mathrm{i}}
\newcommand*{\indic}[1]{\openone_{#1}}

\newcommand*{\R}{\mathbb{R}}
\renewcommand*{\P}{\mathbb{P}}

\newcommand*{\Z}{\mathbb{Z}}
\newcommand*{\N}{\mathbb{N}}
\newcommand*{\ft}[2]{\hat{#1}(#2)}

\newcommand*{\tG}{\tilde{G}}
\newcommand*{\tg}{\tilde{g}}
\newcommand*{\tx}{\tilde{\x}}

\newcommand*{\Dest}{D}
\newcommand*{\tens}[1]{\mathsf{#1}}

\newcommand*{\comp}{\circ}

\newcommand*{\rchan}{\rho^{\textrm{channel}}}
\newcommand*{\rtor}{\rho^{\textrm{torus}}}
\newcommand*{\rbar}{\bar{\rho}}
\newcommand*{\etachan}{\eta^{\textrm{channel}}}
\newcommand*{\etator}{\eta^{\textrm{torus}}}
\newcommand*{\etabar}{\bar{\eta}}

\newcommand*{\hh}[1]{\ft{h}{#1}}
\newcommand*{\fh}[1]{\ft{\phi}{#1}}

\newcommand*{\modulus}[1]{\left| #1 \right|}
\newcommand*{\norm}[1]{\left\| #1 \right\|}
\newcommand*{\supnorm}[1]{\norm{#1}_\infty}
\newcommand*{\fracpart}[1]{\left\{ #1 \right\}}

\newcommand*{\bigO}[1]{\mathcal{O}(#1)}

\renewcommand*{\vec}[1]{\mathbf{#1}}
\newcommand*{\x}{\vec{x}}

\newcommand*{\z}{\vec{z}}

\newcommand*{\B}{\vec{B}}
\renewcommand*{\v}{\vec{v}}

\newcommand*{\vmin}{v_{\text{min}}}
\newcommand*{\wmin}{w_{\text{min}}}

\newcommand*{\normal}[1]{N_{#1}}  % normal dist mean 0,variance 1
\newcommand*{\gauss}[1]{\gamma_{#1}} % gaussian density mean 0 variance 1

\newcommand*{\up}[1]{^{(#1)}}

\begin{document}

\title{Fine structure of distributions and central limit theorem
in diffusive billiards}
\author{David P.\ \surname{Sanders}}
\email{dsanders@maths.warwick.ac.uk}

\affiliation{Mathematics
Institute, University of Warwick, Coventry, CV4 7AL, U.K.}

\begin{abstract}

We investigate deterministic diffusion in periodic billiard
models, in terms of the convergence of rescaled distributions to
the limiting normal distribution required by the central limit
theorem; this is stronger than the usual requirement
that the mean square displacement grow asymptotically linearly in
time. The main model studied is a chaotic Lorentz gas where the
central limit theorem has been rigorously proved. We study
one-dimensional position and displacement densities describing the
time evolution of statistical ensembles in a channel geometry,
using a more refined method than histograms.  We find a
pronounced oscillatory fine structure, and show that this has its
origin in the geometry of the billiard domain. This fine structure
prevents the rescaled densities from converging pointwise to
gaussian densities; however, demodulating them by the fine
structure gives new densities which seem to converge uniformly. We
give an analytical estimate of the rate of convergence of the
original distributions to the limiting normal distribution, based
on the analysis of the fine structure, which agrees well with
simulation results. We show that using a Maxwellian (gaussian)
distribution of velocities in place of unit speed velocities does
not affect the growth of the mean square displacement, but changes
the limiting shape of the distributions to a non-gaussian one.
Using the same methods, we give numerical evidence that a
non-chaotic polygonal channel model also obeys the central limit
theorem, but with a slower convergence rate.

\end{abstract}
\date{\today}

\pacs{05.45.Pq, 02.50.-r, 02.70.Rr, 05.40.Jc}

\maketitle

\section{Introduction} \label{sec:intro}

Diffusion, the process by which concentration gradients are
smoothed out, is one of the most fundamental mechanisms in
physical systems out of equilibrium.  Understanding the
microscopic processes which lead to diffusion on a macroscopic
scale is one of the goals of statistical mechanics
\cite{DorfBook}. Since Einstein's seminal work on Brownian motion
\cite{Gardiner}, diffusion has been modeled by random
processes. However, we expect the microscopic dynamics to be
described by \emph{deterministic} equations of motion.

Recently it has been realized that many simple deterministic
dynamical systems are diffusive in some sense; we call this
\defn{deterministic diffusion}.  Such
systems can be regarded as toy models to understand transport
processes in more realistic systems \cite{DorfBook}.  Examples
include classes of uniformly hyperbolic one-dimensional (1D) maps
(see e.g.\ \cite{KlagesD99} and references therein) and multibaker
models \cite{GaspBook}.  Often rigorous results are not available,
but numerical results and analytical arguments indicate that
diffusion occurs, for example in hamiltonian systems such as the
standard map \cite{LichtenbergLieberman}.

Billiard models, where non-interacting point particles in free
motion undergo elastic collisions with an array of fixed
scatterers, have been particularly studied, since they are
 related to hard sphere fluids, while being
amenable to rigorous analysis \cite{BS, BSC, GaspBook}. They can
also be regarded as the simplest physical systems in which
diffusion, understood as the large-scale transport of mass through
the system, can occur \cite{BunRev}. In this paper we study
deterministic diffusion in two 2D billiard models: a periodic
Lorentz gas, where the scatterers are disjoint disks, and a
polygonal billiard channel.

A definition often used in the physical
literature is that a system is diffusive if the mean square
displacement grows proportionally to time $t$, asymptotically as
$t \to \infty$. However, there are stronger properties which are
also characteristic of diffusion, which a given system may or may
not possess: (i) a \defn{central limit theorem} may be satisfied,
i.e.\ rescaled distributions converge to a normal distribution as
$t \to \infty$;
and (ii) the rescaled dynamics may `look like' Brownian motion.

Two-dimensional (2D) periodic Lorentz gases were proved in \cite{BS,
BSC} to be
diffusive in these stronger senses if they satisfy a geometrical
\defn{finite horizon} condition (\secref{subsec:periodic-lorentz-model}).
We use a square lattice with an additional scatterer in each cell
to satisfy this condition, a geometry previously studied in
\cite{GarrGall, Garrido}. This model is of interest since, unlike
in the commonly studied triangular lattice case (see e.g.\
\cite{GaspBook, MZ, KlagesD00}), we can vary independently two
physically relevant quantities: the available volume in a unit
cell, and the size of its exits; this is possible due to the
two-dimensional parameter space \cite{SandersNext, MyThesis}.

The main focus of this paper is to investigate the fine
structure occurring in the position and displacement distributions
at finite time $t$, and the relation with the convergence to a
limiting normal distribution as $t \to \infty$ proved in \cite{BS, BSC}.
Those papers show in what sense we can
smooth away the fine structure to obtain convergence. However,
from a physical point of view it is important to understand
how this convergence occurs; our analysis provides this.

This analysis makes explicit
 the obstruction that prevents a stronger form of
convergence, showing how density
functions fail to converge pointwise to gaussian densities; it
also allows us to
conjecture a more refined result which takes the fine structure into
account.

Furthermore, this line of argument suggests
 how convergence may occur in other
models where few rigorous results are available.  As an example,
we analyze a recently-introduced polygonal billiard
channel model, showing that the same techniques are still applicable.

\subsection*{Plan of paper} \label{subsec:plan}

In \secref{sec:models} we present the periodic Lorentz gas model for which we
obtain most of our results.
\bsecref{sec:defn-diffusive} discusses the definition of
diffusion in the context of deterministic dynamical systems.
 In \secref{sec:fine-structure}
we study numerically the fine structure of distributions in the
Lorentz gas, finding good agreement with an analytical
calculation in terms of the geometry of the billiard domain, and showing
that when this fine structure is removed, the demodulated densities are
close to gaussian.
This we apply in \secref{sec:clt} to investigate the
central limit theorem and the rate of convergence to the limiting
normal distribution, obtaining a simple estimate of this rate which
agrees well with
numerical results.
In \secref{sec:maxwell-vel-distn} we study the effect of imposing a
Maxwellian (gaussian) velocity distribution in place of a unit speed
distribution, showing that this leads to non-gaussian limiting
distributions.
\bsecref{sec:polyg-bill} extends these ideas to a polygonal
billiard channel, where few rigorous results
are available.
  We finish with conclusions in
\secref{sec:conclusions}.

\section{Two-dimensional periodic Lorentz gas}
\label{sec:models}

%\subsection{Billiard models}

We consider periodic billiard models, where the dynamics can be
studied on the torus. The region $Q$ exterior to the scatterers is
called the \defn{billiard domain}; we denote its area by
$\modulus{Q}$. Since the particles are non-interacting, it is
usual to set all velocities to $1$ by a geometrical rescaling,
although in \secref{sec:maxwell-vel-distn} we discuss the effect
of
 a gaussian velocity distribution.

We focus on a \defn{periodic Lorentz gas}, where the scatterers
are non-overlapping disks. Their strictly convex boundaries make
this a
\defn{scattering} billiard \cite{BS}, and hence a chaotic
system, in the sense that it has a positive Lyapunov exponent
\cite{CherMark, GaspBook} and positive Kolmogorov--Sinai entropy
\cite{GaspBook}.

\subsection{Periodic Lorentz gas model}
\label{subsec:periodic-lorentz-model}

\begin{figure}
\subfigure[]{\label{fig:2dlattice}
\includegraphics{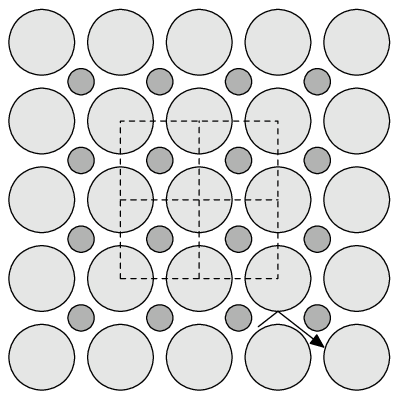}}
\hfill \subfigure[]{\label{fig:geom2d}
\includegraphics{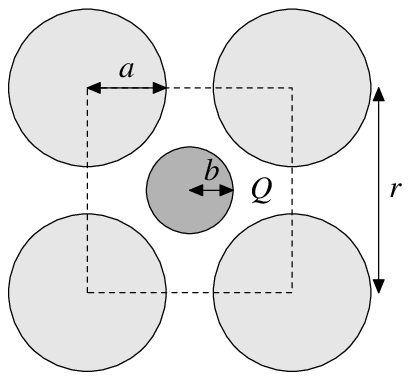}}
\caption{\label{fig:2d-geometry} (a) Part of the infinite system,
constructed from two square lattices of disks shown in different
shades of gray; dashed lines indicate several unit cells and an elastic
collision is shown. (b) A
single unit cell, defining the geometrical parameters.  The
billiard domain is the area $Q$ exterior to the disks.}
\end{figure}

The model we study, previously considered in
\cite{GarrGall,Garrido}, consists of two square lattices of disks;
they have the same lattice spacing $r$, and radii $a$ and $b$,
respectively, and are positioned such that there is a $b$-disk at
the center of each unit cell of the $a$-lattice: see
\figref{fig:2d-geometry}. In analytical calculations we take the
length scale as $r=1$, as in \cite{Garrido, GarrGall}, whereas in
numerical simulations we  fix $a=1$ and scale $r$ and $b$
appropriately, as in \cite{KlagesD00}.

\paragraph*{Finite horizon condition}
Periodic Lorentz gases were shown in \cite{BS, BSC} to be
diffusive
 (\secref{sec:defn-diffusive}), provided they satisfy the
\defn{finite horizon} condition:
  there is an upper bound on the free path length between collisions.
 If this is not the case, so that a particle can travel infinitely far
without colliding (the billiard has an
\defn{infinite horizon}), then \defn{corridors} exist \cite{Bleher}, which
allow for fast propagating trajectories, leading to super-diffusive
behavior, as was recently rigorously proved \cite{SzaszVarjuII}.

We restrict attention to parameter values within the finite
horizon regime by choosing $b$ to block all  corridors
\cite{Garrido, SandersNext}.

\subsection{Statistical properties} \label{subsec:stat-props}
Statistical properties of deterministic dynamical systems arise
from an ensemble of initial conditions $(\x_0,\v_0)$ modeling the
imprecision of physical measurements. We always take a uniform distribution
with respect to
\defn{Liouville measure} in one unit cell: the positions $\x_0$ are uniform
with respect to Lebesgue measure in the billiard domain $Q$,
and the velocities $\v_0$ are uniform in the unit circle $S^1$,
i.e.\ with angles between $0$ and $2 \pi$, and unit speeds.

We evolve $(\x_0, \v_0)$ for a time $t$ under the billiard  flow
$\Phi^t$ in phase space
 to $(\x(t), \v(t))$.  Note that
Liouville measure on the torus is invariant under this flow
\cite{CherMark}.  In numerical experiments, we take a large sample
$(\x\up{i}_0,\v\up{i}_0)_{i=1}^N$ of size $N$ of initial
conditions chosen uniformly with respect to Liouville measure
using a random number generator. These evolve after time $t$ to
$(\x\up{i}(t),\v\up{i}(t))_{i=1}^N$; the distribution of this
ensemble then gives an approximation to that of $(\x(t), \v(t))$.

We denote averages over the initial conditions, or
equivalently expectations with respect to the distribution of
$(\x_0, \v_0)$, by $\langle \cdot
\rangle$.  Approximations of such averages can be evaluated using a simple
Monte Carlo
method \cite{NR} as
\begin{equation}\label{}
\mean{f(\x_0,\v_0)} = \lim_{N \to \infty} \frac{1}{N}\sum_{i=1}^N
f(\x_0\up{i},\v_0\up{i}).
\end{equation}
The infinite sample size limit, although unobtainable in practice,
reflects the expectation that larger $N$ will give a better
approximation.  Averages at time $t$ can be evaluated  by using a
function $f$ involving $\flow^t$.

\subsection{Channel geometry} \label{subsec:channel-geometry}

Diffusion occurs in the extended system obtained by  unfolding the
torus to a 2D infinite lattice: see \cite{BS, BSC} and
\secref{sec:defn-diffusive}. The diffusion process is then
described by a second order diffusion tensor having
 $4$ components $D_{ij}$ with respect to a given orthonormal
basis, given by
\begin{equation}\label{}
D_{ij} = \lim_{t \to \infty} \frac{1}{2t} \mean{\Dx_i \Dx_j}_t.
\end{equation}

\begin{figure}
 \includegraphics{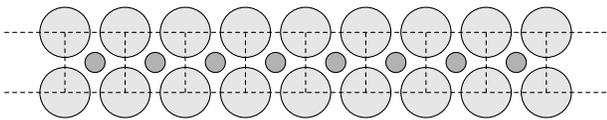}
  \caption{1D channel obtained by unfolding torus
in $x$-direction.}
  \label{fig:1dlattice}
\end{figure}

The square symmetry of our model reduces the diffusion tensor to a
constant multiple $D$ of the identity tensor; we can evaluate this
\defn{diffusion coefficient} by restricting attention to the dynamics in a
$1$-dimensional \defn{channel} extended only in the $x$-direction; see
\figref{fig:1dlattice}.  Correspondingly, we restrict attention to
1D marginal distributions.

\begin{figure}
\subfigure[]{\label{fig:lorentz-channel}
\includegraphics{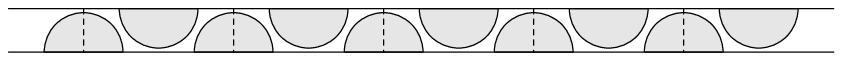}}
\hfill
\subfigure[]{\label{fig:unfolded-lorentz-channel}
\includegraphics{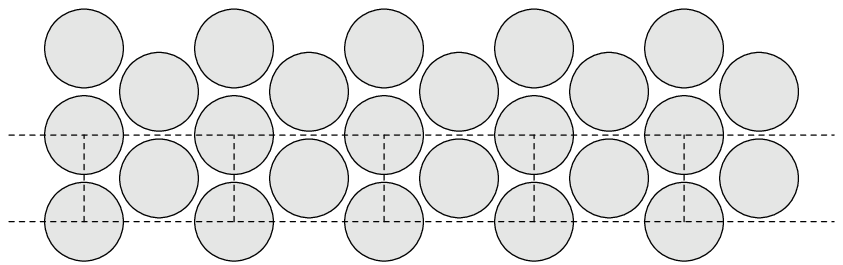}}
\caption{
(a) Lorentz channel studied in \cite{GaspChaotScattering,
  AlonsoLorentzChannel} with hard upper and lower boundaries;
  dotted lines indicate unit cells.
(b) Fully unfolded triangular Lorentz gas.
  Dotted lines indicate unit cells forming a channel with periodic upper and
  lower boundaries.}
\end{figure}

A channel geometry, with hard horizontal boundaries, corresponding
to the triangular Lorentz gas was studied in
\cite{GaspChaotScattering, AlonsoLorentzChannel}
(\figref{fig:lorentz-channel}).
  This is equivalent to
 a channel with twice the original height and
\emph{periodic} boundaries, shown in
\figref{fig:unfolded-lorentz-channel} as part of the whole
triangular lattice obtained by unfolding completely in the
vertical direction. We can view this  lattice as consisting of
rectangular unit cells (\figref{fig:unfolded-lorentz-channel})
which are
 stretched versions of the square unit cell considered
above, with the extra condition $a=b$. The results in the
remainder of this paper then extend to this case with minor
changes.

\section{Deterministic diffusion}
\label{sec:defn-diffusive}

In this section we briefly recall how to make precise the fact that the
behavior of certain deterministic dynamical systems `looks like' that of
the diffusion equation.

\subsection{Diffusion as a stochastic process}
\label{subsec:diffn-as-stoch-proc}
 Diffusion is described classically by the diffusion
equation
\begin{equation}\label{eq:diffn-eqn}
\pd{\rho(t, \x)}{t} = D \,   \nabla^2 \rho(t, \x),
\end{equation}
where $\rho$ is the density of the diffusing substance. Following
Einstein and Wiener (see e.g.\ \cite{Gardiner}), we can model
diffusion as a stochastic process $B_t$, determined by the
probability density $p(\x,t)$ of a particle being at position $\x$
at time $t$ given that it started at $\x=\vec{0}$ at time $t=0$.

Imposing conditions on the process determined from physical
requirements gives a \emph{diffusion process}, where $p(\x,t)$
satisfies the equation
\begin{equation}\label{eq:Fokker-Planck}
\pd{p}{t}  + \pd{}{x_i} \left[A_i
\, p - \frac{1}{2} \sum_j \pd{}{x_j} \left(B_{ij}\, p \right) \right] = 0,
\end{equation}
known as \defn{Kolmogorov's forward equation} or the
\defn{Fokker--Planck equation}
\cite{Gardiner}. The  \defn{drift vector} $\vec{A}(\x,t)$ and the
\defn{diffusion tensor} $\tens{B}(\x,t)$
  give the mean and variance, respectively, of  infinitesimal displacements
at position $\x$ and time $t$ \cite{Gardiner}.

If the system is sufficiently symmetric that the drift
is zero and the diffusion tensor
is a multiple of the identity tensor, then the  process is
\defn{Brownian motion},
and \eqref{eq:Fokker-Planck} reduces to the diffusion equation
\eqref{eq:diffn-eqn}.   A general diffusion process, however, can
be
\defn{inhomogeneous} in both space and time.

\subsection{Diffusion in dynamical systems via limit theorems}
\label{subsec:DS-prob-limit-thms} Diffusion in billiards concerns
the statistical behavior of the particle positions.  We can write
the first component $x_t$ of the position $\x_t$ at time $t$ as
\begin{equation}\label{eq:x-as-integral}
x_t = \int_0^t v_1(s) \rd s + x_0 =
\int_0^t f \comp \Phi^s(\cdot) \rd s + x_0 ,
\end{equation}
where $f = v_1$, the first velocity component.  This expresses $x_t$ solely in
terms of functions defined on the torus.  In fact, \eqref{eq:x-as-integral}
shows that
the displacement $\Dx_t \defeq x_t-x_0$ is in some sense a more natural
observable than the position $x_t$ in this context.

We thus wish to study the distribution of accumulation functions
of the form  $S_t(\cdot) \defeq \int_0^t f \comp \Phi^s(\cdot) \rd
s$, in particular in  the limit as $t \to \infty$
\cite{CherYoung}. We remark that other observables $f$ are
relevant for different transport processes \cite{BunRev}.

We denote by $\flow^t \from \M \to \M$ the flow of a
 dynamical system with time
$t \in \R$. Given a probability measure $\mu$ describing the
distribution of initial conditions, we can find the probability of
being in certain regions of the phase space $\M$ at given times,
so that we have a stochastic process. If the measure $\mu$ is
\defn{invariant}, so that $\mu(\flow^{-t}(A)) = \mu(A)$ for all
times $t$ and all nice sets $A$, then the stochastic process is
\defn{stationary} \cite{CherYoung}.

The integral in the definition of $S$ is then a continuous-time
version of a Birkhoff sum $\sum_{i=0}^{n-1} f \comp \Phi^i$ over
the stationary stochastic process given by $\Phi$, so that we may
be able to apply
 limit theorems  from the theory of stationary stochastic
processes \cite{CherYoung}.
For the case of the periodic Lorentz gas with finite horizon,
it was proved in \cite{BS, BSC}
that the
following limit theorems hold.

\paragraph{Asymptotic linearity of mean square displacement}

 The limit
\begin{equation}\label{}
2D \defeq \lim_{t\to\infty} \frac{1}{t} \mean{\Dx^2}_t
\end{equation}
exists, so that the mean square displacement  $\meanat{\Dx^2}{t} \defeq
\mean{\left[\Dx(t)\right]^2}$
(the variance of the displacement distribution)
grows asymptotically linearly in time:
\begin{equation}\label{}
\meanat{\Dx^2}{t} \sim 2Dt \quad \text{as }t \to \infty,
\end{equation}
where $D$ is the \defn{diffusion coefficient}.
In $d \ge 2$ dimensions, setting $\Dx_i(t) \defeq x_i(t) - x_i(0)$, we have
\begin{equation}\label{}
\meanat{\Dx_i \, \Dx_j}{t} \sim 2 D_{ij} t,
\end{equation}
where the $D_{ij}$ are components of a symmetric diffusion
tensor.

\paragraph{Central limit theorem: convergence to normal distribution}
Scale the displacement distribution by $\sqrt{t}$, so that the variance of the
rescaled distribution is bounded. Then this distribution converges
\defn{weakly}, or \defn{in distribution}, to a normally
distributed random variable $\z$ \cite{GaspNic, CherYoung}:
\begin{equation}\label{}
    \frac{\x(t)-\x(0)}{\sqrt{t}} \distconv
  \z, \qquad \text{as } t \to \infty.
\end{equation}
In the $1$-dimensional case, this means that
\begin{equation}\label{eq:CLT-defn}
\lim_{t \to \infty} \P\left(\frac{x_t-x_0}{\sqrt{t}} < u\right)
= \frac{1}{\sigma \sqrt{2 \pi}} \int_{s=-\infty}^u
\e^{-s^2/2\sigma^2} \rd s,
\end{equation}
where $\P(\cdot)$ denotes probability with respect to the
distribution of the initial conditions, and $\sigma^2$ is the
variance of the limiting normal distribution. In $d \ge 2$
dimensions, this is replaced by similar statements about
probabilities of $d$-dimensional sets. This is the
\defn{central limit theorem} for the random variable $\Dx$.
From (a) we know that in 1D, the variance of the limiting normal
distribution is $\sigma^2 = 2D$; in
$d \ge 2$ dimensions, the covariance matrix of $\z$ is given by
the matrix $(2 D_{ij})$ \cite{BS, DettCoh2}.

\paragraph{Functional central limit theorem: convergence of path
distribution to Brownian motion}

We rescale the  path $\x_t$ by the scale from (b), defining
$\tx_t$ by \cite{Bleher}
\begin{equation}\label{}
\tx_t(s) \defeq \frac{\x(st)-\x(0)}{\sqrt{t}}, \quad s \in [0,1].
\end{equation}
The distribution of  these rescaled paths then converges
in distribution
to Brownian motion:
\begin{equation}
\tx_t \distconv \B \quad \text{as } t \to \infty,
\end{equation}
where the Brownian motion $\B$ has covariance matrix as in (b).
 This is known as a \defn{functional central limit theorem}, or
\defn{weak invariance principle} \cite{CherYoung}.

A sufficient condition for this is  that the following two
properties hold \cite{Billingsley}.
(i) The \defn{multi-dimensional central limit theorem},
a generalization of (b), is satisfied. This says that the
 finite-dimensional distributions of the process
$\tx_t$ converge to those of Brownian motion, so that
for any $n$, any times $s_1 < \cdots <
s_n$, and any reasonable sets $D_1, \ldots, D_n$ in $\R^d$, we
have
\begin{multline}\label{eq:defn-fclt}
\prob{\tx_t(s_1) \in D_1, \ldots, \tx_t(s_n) \in D_n} \\
\stackrel{t \to \infty}{\longrightarrow} \prob{\B(s_1) \in D_1,
\ldots, \B(s_n) \in D_n}.
\end{multline}
The right-hand side can be expressed as a multi-dimensional
integral over gaussians: see e.g.\ \cite{DettCoh2, MyThesis}. (ii)
The convergence is \defn{tight}, which prevents mass escaping to
infinity: see \cite{Billingsley} for the definition.

\subsection{Discussion of definitions of diffusion}

Property (c) is the strongest sense in which a dynamical system
can show
 deterministic diffusion,
 making precise how a rescaled dynamical system can look like Brownian
 motion.
However, few physically relevant systems have been proved to
satisfy (c): interest in the periodic Lorentz gas comes largely
from the fact that it is one; another is the triple linkage
\cite{HuntMackay}.

The multi-dimensional central limit theorem part
of (c) was studied in \cite{DettCoh2}, where both
Lorentz gases and wind--tree models
were found to obey it, tested for certain sets $D_i$ and certain
values of $n$.  However, as stated in \cite{DettCoh2}, (c) is difficult to
investigate numerically, and the results in that paper seem to be the
best that we can expect.

Property (b), the central limit theorem, has been shown for large
classes of observables $f$ in many dynamical systems (see
\cite{CherYoung} and references therein),
 but
again they are often not physical.  Property (b) was used in
\cite{GaspNic} as the definition of a diffusive system, but does
not seem to have been applied in the physical literature; it is
the approach taken in this paper.

Many papers in the physical literature define a system to be
diffusive if only property (a) is verified (numerically), e.g.\
\cite{KlagesD00,AlonsoPolyg,DettCoh1}.  Many types of system are
diffusive in this sense, including 1D maps \cite{KlagesD99},
random Lorentz gases \cite{DettCoh1} and Ehrenfest wind--tree
models, both periodic \cite{AlonsoPolyg} and random
\cite{DettCoh1}.

It is possible for the weaker properties to hold when the stronger
ones do not. For example, in \cite{CohenKong} a \emph{disordered}
lattice-gas wind--tree model was reported to have an
asymptotically linear mean square displacement, but a non-gaussian
distribution function, i.e.\ (a) but not (b). However, disorder
can lead to trapping effects which cannot occur in periodic
systems \cite{AlonsoPolyg}, and we are not aware of a
\emph{periodic} (and hence ordered) billiard-type model with
unit-speed velocity distribution which shows (a) but not (b),
although in \secref{sec:maxwell-vel-distn} we show that this can
occur with a Maxwellian velocity distribution.

\section{Fine structure of position and displacement
distributions} \label{sec:fine-structure}

We now focus on the diffusive properties of the periodic Lorentz
gas model introduced in \secref{sec:models}.  In this section, we
describe the fine structure of position and displacement distributions.
The displacement distribution occurs naturally in the central
limit theorem (\secref{subsec:DS-prob-limit-thms}) and in
Green--Kubo relations \cite{DorfBook, GaspBook}, whereas the
position distribution is more natural if we are unable to track
the paths of individual particles.  It is possible to show that
the asymptotic properties of the position and displacement distributions
are the same, in the sense that one has an asymptotically linear growth if
and only if the other does, and similarly for the central limit theorem
\cite{MyThesis}.
It is hence equivalent to consider diffusive properties by
studying either distribution.

\subsection{Position and displacement distributions}
\label{subsec:posn-disp-distns}

\begin{figure*}
\includegraphics{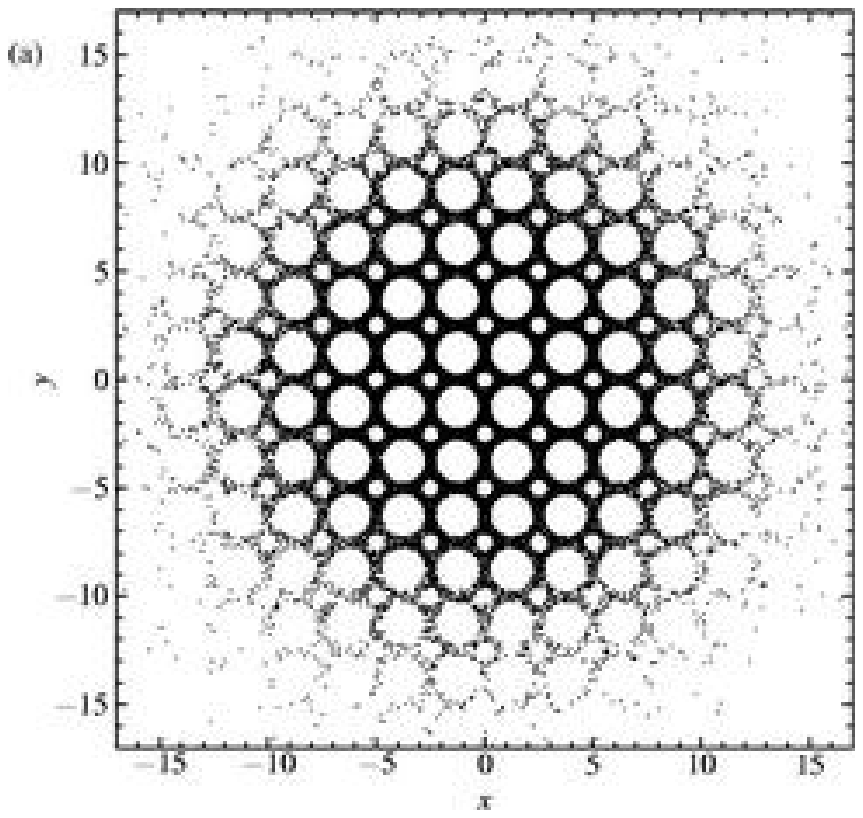}
\hfill
\includegraphics{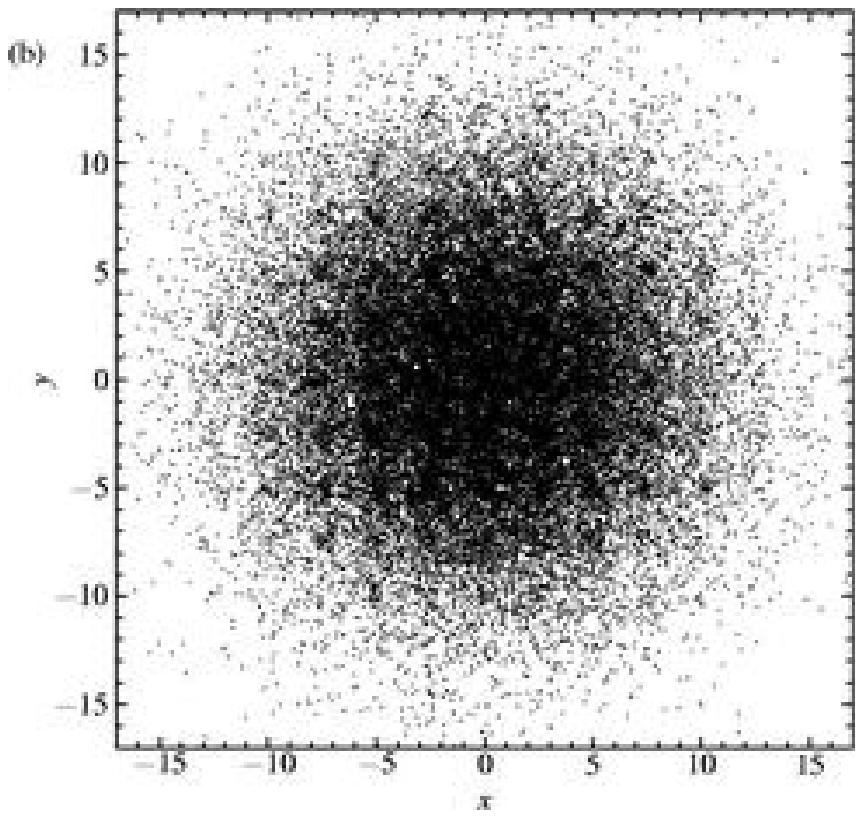}
\caption{\label{fig:2d-distns}
(a) 2D position distribution; (b) 2D displacement
distribution. $r=2.5$; $b=0.4$; $t=50$; $N=5 \times 10^4$ initial conditions.
}
\end{figure*}

\bfigref{fig:2d-distns} shows scatterplots representing 2D
position and displacement distributions for a representative
choice of geometrical parameters. Each dot represents one initial
condition started in the central unit cell and evolved for time
$t=50$; $N=5 \times 10^4$ samples are shown.  Both distributions
show decay away from a maximum in the central cell, an overall
circular shape, and the occurrence of a periodic fine structure.

These figures are projections to the billiard domain $Q$ of the
density in the phase space $Q \times S^1$. Since the dynamics on
the torus is mixing \cite{CherMark}, the phase space density
converges \emph{weakly} \cite{LasotaMackey} to a uniform density
on phase space
 corresponding to the invariant Liouville
measure. Physically, the phase space density develops a
complicated layer structure in the stable direction of the
dynamics: see e.g.\ \cite{DorfBook}.  Projecting corresponds to
integrating over the velocities; we expect this to eliminate this
complicated structure and result in some degree of smoothness of
the projected densities.  However, we are not aware of any
rigorous results in
 this direction, even
for relatively well-understood systems such as the Arnold cat map
\cite{DorfBook}.

These 2D distributions are difficult to work with, and we instead
restrict attention to one-dimensional marginal distributions,
i.e.\ projections onto the $x$-axis, which will also have some
degree of smoothness.
 We denote the 1D position
density at time $t$ and position $x \in \R$ by $f_t(x)$
and the displacement density for displacement $x$ by $g_t(x)$.
We let
their respective (cumulative) distribution functions be $F_t(x)$ and
$G_t(x)$, respectively, so that
\begin{equation}\label{}
F_t(x) \defeq \P(x_t \le x) = \int_{-\infty}^x f_t(s) \rd s,\\
\end{equation}
and similarly for $G_t$.  (When necessary, we will instead denote
displacements by $\xi$.)
The densities show the
structure of the distributions more clearly, while the distribution
functions are
more directly related to
analytical considerations.

\subsection{Numerical estimation of  distribution
functions and densities} \label{subsec:num-estimation-densities}

We wish to estimate numerically the above denstities and
distribution functions at time $t$ from the $N$  data points
$x_t\up{1}, \ldots, x_t\up{N}$. The most widely used method in the
physics community for estimating density functions from numerical
data  is the histogram; see e.g.\ \cite{AlonsoPolyg}. However,
histograms are not always appropriate, due to their non-smoothness
and dependence on bin width and position of bin origin
\cite{Silverman}. In \cite{AlonsoPolyg}, for example, the choice
of a coarse  bin width obscured the fine structure of the
distributions that we describe in \secref{sec:polyg-bill}.

We have chosen the following alternative method, which seems to
work well in our situation, since it is able to deal with strongly
peaked densities more easily, although we do not have any rigorous
results to justify it. We have also checked that histograms and
kernel density estimates (a generalization of the histogram
\cite{Silverman}) give similar results, provided sufficient care
is taken with bin widths.

We first calculate the empirical cumulative distribution function
\cite{Scott, Silverman}, defined by $F_t^{\emp}(x) \defeq
\#\{i:x_t\up{i} \le x\}$ for the position distribution, and
analogously for the displacement distribution.
The estimator $F_t\emp$ is the optimal one for the distribution
function $F_t$ given the data, in the sense that there are no
other unbiased estimators with smaller variance \cite[p.\
34]{Scott}. We find that the distribution functions in our models
are smooth on a scale larger that that of individual data points,
where statistical noise dominates. (Here we use `smooth' in a
visual, nontechnical sense; this corresponds to some degree of
differentiability). We verify
that adding more data does not qualitatively change this
larger-scale structure: with $N=10^7$ samples we
seem to capture the fine structure.

We now wish to construct the density function $f_t = \partial F_t
/ \partial x$.  Since the direct numerical derivative of $F_t\emp$ is
useless due to
statistical noise, our procedure is
to fit an (interpolating)
\defn{cubic spline} to an evenly-spread sample of points from $F_t\emp$,
and differentiate the cubic
spline to obtain the density function at as many points as required
\cite{MyThesis}.  Sampling evenly from $F_t\emp$
automatically uses more samples where the data are more highly concentrated,
i.e.\ where the
density is larger.

 We must confirm
(visually or in a suitable norm) that our spline approximation
reproduces the fine structure of the distribution function
sufficiently well, whilst ignoring the variation due to noise on a
very small scale. As with any density estimation method, we have
thus made an assumption of smoothness \cite{Silverman}. The
analysis of the fine structure in \secref{sec:fine-structure}
justifies this to some extent.

\subsection{Time evolution of 1D distributions}
\label{subsec:time-evolution-distns}

\begin{figure*}
\includegraphics{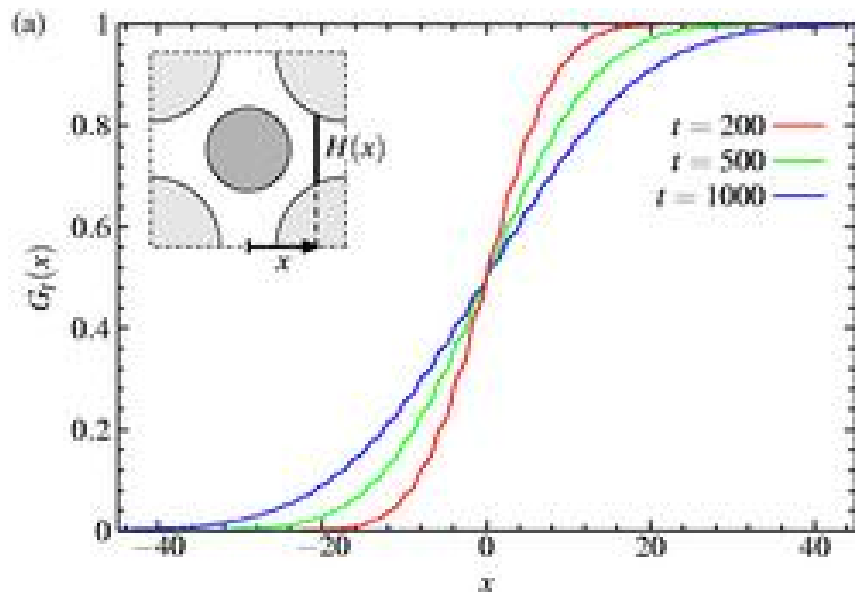}
\hfill
\includegraphics{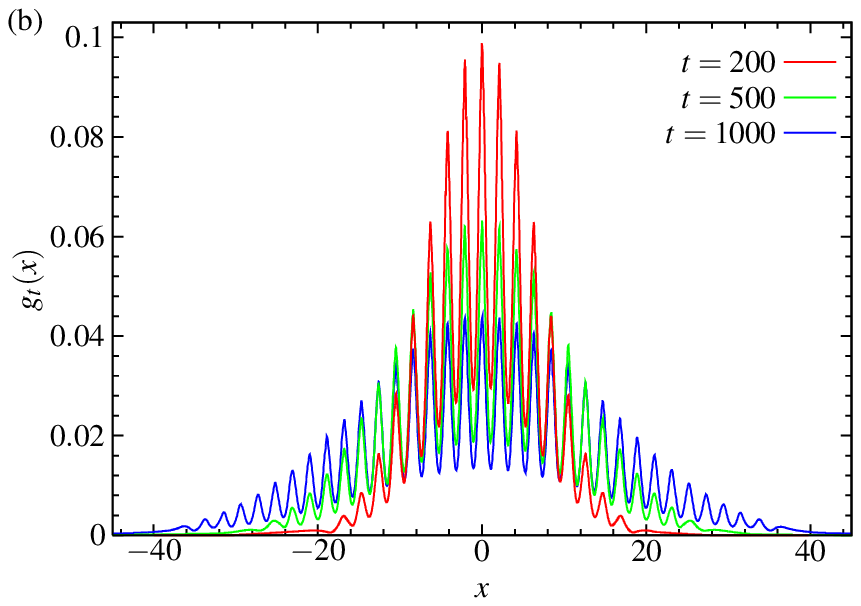}
\caption{\label{fig:time-evol}\label{fig:H-of-x}(Color online)
(a) Time evolution of displacement distribution
functions. (b) Time evolution of displacement densities,
calculated by numerically
  differentiating a cubic spline approximation to distribution functions.
$r=2.1$; $b=0.2$.  The inset in (a) shows the definition of the set
$H(x)$ required later.}
\end{figure*}

\bfigref{fig:time-evol} shows the time evolution of 1D
displacement distribution functions and densities for certain
geometrical parameters, chosen to emphasize the oscillatory
structure. Other parameters within the finite horizon regime give
qualitatively similar behavior.

The distribution functions are smooth, but have a
step-like structure. Differentiating the spline
approximations to these distribution functions gives
 densities which have an
underlying gaussian-like shape, modulated by a \emph{pronounced
fine structure} which persists at all times (\figref{fig:time-evol}(b)).
This fine structure
is just noticeable in Figs.\ 4 and 5 of \cite{AlonsoPolyg}, but
otherwise does not seem to have been reported previously, although
in the context of iterated 1D maps a  fine structure
was found, the origin of which is pruning effects: see
 e.g.\ Fig.\ 3.1 of
\cite{KlagesPhD}.  We will show
that in billiards this fine structure can be understood
by considering the
geometry of the billiard domain.

\subsection{Fine structure of position density}
\label{subsec:unfold-posn-density}

Since Liouville measure on the torus is invariant, if
the initial distribution is uniform with respect to
Liouville measure, then the distribution at any time $t$ is still uniform.
Integrating over the velocities, the position distribution at time
$t$ is hence always uniform with respect to Lebesgue measure in
the billiard domain $Q$, which
 we normalize such that the measure of $Q$ is $1$.
Denote the two-dimensional position density on the torus at
$(x,y) \in [0,1)^2$ by
$\rtor(x,y)$.  Then
\begin{equation}\label{}
\rtor(x,y) = \frac{1}{\modulus{Q}} \indic{Q}(x,y) =
\frac{1}{\modulus{Q}} \indic{H(x)}(y).
\end{equation}
Here, $H(x) \defeq \{y: (x,y) \in Q\}$ is the
set of allowed $y$ values for particles with horizontal coordinate
$x$ (\figref{fig:H-of-x}(a) inset),
%$\modulus{Q}$ is the area of $Q$,
and $\indic{B}$ is the indicator function of the (one- or
two-dimensional) set $B$, given by
\begin{equation}\label{}
\indic{B}(b) = \begin{cases} 1,\quad \text{if } b \in B \\
0, \quad \text{otherwise}.
\end{cases}
\end{equation}
Thus for fixed $x$, $\rtor(x,y)$ is
independent of $y$ within the available space $H(x)$.

Now unfold the dynamics onto a 1-dimensional channel in the
$x$-direction, as in \figref{fig:1dlattice}, and consider the torus as the
distinguished unit cell at the origin. Fix a vertical line with
horizontal coordinate $x$
in this cell,
 and consider its periodic translates $x+n$ along the channel,
where $n \in \Z$. Denoting the density there by $\rchan_t(x+n, y)$,
we  have that for all $t$ and for all $x$ and $y$,
\begin{equation}\label{eq:reduce_density_torus}
 \sum_{n \in \Z} \rchan_t(x+n, y) = \rtor(x,y).
\end{equation}

We expect that after a sufficiently long time,
the distribution within cell  $n$  will look
like the distribution on the torus, modulated by a slowly-varying
function of $x$.
In particular, we expect that the 2D position density will become
asymptotically
uniform in $y$ within
 $H(x)$ at long times.
We have not been able to prove this, but we have checked by
constructing
2D kernel density estimates \cite{Silverman} that it seems to be correct.
A `sufficiently long'
time would be one which is much longer than the time required for the
diffusion process to
cross one unit cell.

Thus we have approximately
\begin{equation}\label{eq:first-def-rho-bar}
\rchan_t(x,y) \simeq \rtor(x,y) \, \rbar_t(x) = \rbar_t(x)
\frac{1}{\modulus{Q}} \indic{H(x)}(y),
\end{equation}
where $\rbar_t(x)$ is the
\defn{shape} of
the two-dimensional density distribution as a function of $x \in
\R$; we expect this to be a slowly-varying function. We use
`$\simeq$' to denote that this relationship holds in the long-time
limit, for values of $x$ which do not lie in the tails of the
distribution. Although this breaks down in the tails,
 the density is in any case small there.

The 1D marginal density that we measure will then be given approximately by
\begin{equation}\label{eq:1d-posn-density-from-2d}
f_t(x) = \int_{y=0}^1 \rchan_t(x,y) \rd y \simeq \rbar_t(x) \,
h(x),
\end{equation}
where $h(x) \defeq \modulus{H(x)}/\modulus{Q}$ is the normalized
height (Lebesgue measure) of the set $H(x)$ at position $x$ (see
the inset of \figref{fig:H-of-x}(a)).  Note that $H(x)$ is not
necessarily a connected set.

 Thus the measured density $f_t(x)$ is
given by the shape $\rbar_t(x)$ of the 2D density,
\defn{modulated} by  fine-scale oscillations due to the geometry
of the lattice and described by $h(x)$, which we call the
\defn{fine structure function}.

The above argument motivates the \emph{(re-)definition} of $\rbar_t(x)$
so that
that $f_t(x) =  h(x) \rbar_t(x)$, now with strict equality and for all times.
We can then view $\rbar_t(x)$ as the density with respect to
a \emph{new underlying measure} $h \, \lambda$, where $\lambda$ is
$1$-dimensional Lebesgue measure;
this measure takes into account the available space, and is hence more
natural in this
problem.  We expect that $\rbar_t$ will now describe the large-scale shape
of the density, at least for long times and $x$ comparatively small.

%\subsection{Numerical demodulation of position density}

\begin{figure}
\includegraphics{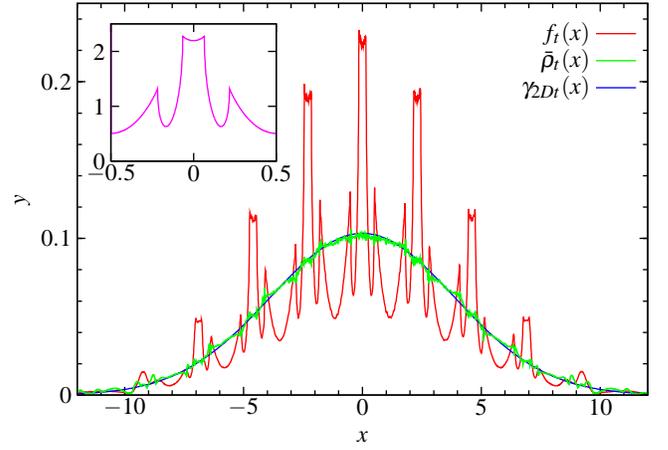}
\caption{\label{fig:demodulate-posn-distn}
(Color online) Position density $f_t$ exhibiting a pronounced fine structure,
together with the demodulated slowly-varying function $\rbar_t$
and a gaussian with variance $2Dt$. The inset shows one period of the
demodulating fine structure function $h$. $r=2.3$;
$b=0.5$; $t=50$.}
\end{figure}

\bfigref{fig:demodulate-posn-distn} shows the original and demodulated
densities $f_t$ and
$\rbar_t$ for a representative choice of geometrical parameters.  The
fine structure in $f_t$ is very pronounced, but is eliminated
nearly completely when demodulated by dividing by the fine structure $h$,
leaving
a demodulated density $\rbar_t$ which is close to
the gaussian density with variance $2Dt$ (also shown).

We estimated the diffusion coefficient $D$ as follows.
For $r=2.3$ and $b=0.5$, using $N=10^7$ particles evolved to
$t=1000$, the best fit line for $\log \msd_t$ against $\log t$ in the
region $t \in [500,1000]$ gives $\msd \sim t^{1.00003}$, which we
regard as confirmation of asymptotic linear growth. Following
\cite{KlagesD00}, we use the slope of $\log \msd_t$ against $t$ in
that region as an estimate of $2D$, giving $D = 0.1494 \pm 0.0002$;
see \cite{SandersNext, MyThesis} for
the error analysis.

(Throughout the paper, we denote by $\gauss{\sigma^2}$
the gaussian density with mean $0$ and variance $\sigma^2$, and by
 $\normal{\sigma^2}$ the corresponding normal distribution function.)

 Note that although the density has non-smooth points, which affects the
 smoothness assumption in our density estimation procedure described in
 \secref{subsec:num-estimation-densities}, in practice these points are
 still handled reasonably well.  If necessary, we could treat these points
     more carefully, by suitable choices of partition points in that method.

\subsection{Fine structure of displacement density}
\label{subsec:unfold-disp-density}

We can treat the displacement density similarly, as follows.
Let $\eta_t(x,y)$ be the 2D displacement density function at time
$t$, so that
\begin{equation}\label{}
\int_{-\infty}^x \int_{\infty}^y \eta_t(x,y) \rd x \rd y =
\prob{\D x_t \le x, \D y_t \le y}, \quad x,y \in \R.
\end{equation}
(Recall that $\D x_t \defeq x_t - x_0$.)
 We
\emph{define} the projected versions $\etachan$ and $\etator$ as
follows:
\begin{gather}\label{}
\etachan_t(x,y) \defeq \sum_{n \in \Z} \eta_t(x, y+n),
\quad x \in \R, y \in [0,1), \\
\etator_t(x,y) \defeq \sum_{n \in \Z} \etachan_t(x+n, y), \quad
x,y \in [0,1).
\end{gather}
Again we view the torus as the unit cell at the origin where all
initial conditions are placed.
Note that
 projecting the displacement distribution on $\R^2$
 to the channel or torus gives the same result as
first
 projecting and then obtaining the
displacement distribution in the reduced geometry.
Hence the designations as being associated with the
channel or torus are appropriate.

Unlike $\rtor$ in the previous section, $\etator_t$ is not
independent of $t$: for example, for small enough $t$, all displacements
 increase with time.  However, we show that $\etator_t$ rapidly approaches a
distribution which \emph{is} stationary in time.

Consider a small ball of initial conditions of positive Liouville
measure around a point $(\x,\v)$.  Since the system is mixing on
the torus, the position distribution at time $t$ corresponding to
those initial conditions converges as $t \to \infty$ to a
distribution which is uniform with respect to Lebesgue measure in
the billiard domain $Q$. The
 corresponding limiting displacement distribution is hence obtained by
averaging the displacement of
$\x$ from all points on the torus.

Extending this to an initial distribution which is uniform with
respect to Liouville measure over the whole phase space, we see
that the limiting displacement distribution is given by averaging
displacements of two points in $Q$, with both points distributed
uniformly with respect to Lebesgue measure on $Q$. This limiting
distribution we denote by $\etator(x,y)$, with no $t$ subscript.
%We expect, and have checked numerically, that the approach of the
%displacement distribution $\etator_t$  to the limiting
%distribution $\etator$ is fast.

As in the previous section, we expect the $y$-dependence of
$\etachan_t(x+n,\cdot)$ to be the same, for large enough $t$, as
that of $\etator(x,\cdot)$ for $x \in [0,1)$. However,
$\etator(x,\cdot)$ is not independent of $y$, as can be seen from
a projected version of \figref{fig:2d-distns}(b) on the torus
\cite{MyThesis}. We thus set
\begin{equation}\label{eq:def-eta-chan}
\etachan_t(x,y) \simeq \etator(x,y) \, \etabar_t(x).
\end{equation}

To obtain the 1D marginal density $g_t(x)$, we
integrate with respect to $y$:
\begin{equation}\label{}
g_t(x) = \int_{y=0}^1 \etachan_t(x,y) \rd y \simeq \phi(x)
\etabar_t(x),
\end{equation}
where
\begin{equation}\label{}
\phi(x) \defeq \int_{y=0}^1 \etator(x,y) \rd y.
\end{equation}
Again we now redefine $\etabar$ so that $g_t(x) = \phi(x)
\etabar_t(x)$, with the
 fine structure of $g_t(x)$ being described by $\phi$ and the
large-scale variation by $\etabar(x)$,  which can be regarded as the
density with respect to the new measure
$\phi \, \lambda$
taking account of the excluded volume.
In the next section we
evaluate $\phi(x)$ explicitly.

\subsection{Calculation of $x$-displacement density $\phi(x)$ on torus }
\label{subsec:x-disp-density-torus}

Let $(X_1, Y_1)$ and $(X_2, Y_2)$ be independent
random variables, distributed uniformly with respect to Lebesgue
measure in the billiard domain $Q$, and let $\DX \defeq
\fracpart{X_2-X_1} \in [0,1)$ be their
$x$-displacement
(where $\fracpart{\cdot}$ again denotes the fractional part of its argument).
Then
$\DX$ is the sum of two independent random variables, so that  its
density  $\phi$ is given by the following convolution, which
correctly takes account of the periodicity of $h$ and $\phi$ with
period $1$:
\begin{equation} \label{eq:convolution}
  \phi(\xi) = \int_0^1 h(x) \, h(x+\xi) \rd x.
\end{equation}
This form leads us to expand in
Fourier series:
\begin{equation}\label{eq:fourierseries}
h(x) =  \sum_{k \in \Z} \hh{k} \, \e^{2 \pi \i k x} = \hh{0} + 2
\sum_{k \in \N} \hh{k} \, \cos 2 \pi k x,
\end{equation}
and similarly for $\phi$, where the Fourier coefficients are
defined by
\begin{equation}\label{eq:fouriercoeffs}
  \hh{k} \defeq \int_0^1 h(x) \, \e^{-2 \pi \i k x} \rd x
          =   \int_0^1 h(x) \, \cos(2 \pi k x) \rd x.
\end{equation}
 The last equality follows from the evenness of $h$, and shows
 that $\hh{k} = \hh{-k}$, from which the second
 equality in \eqref{eq:fourierseries} follows.
 Fourier transforming \eqref{eq:convolution} then gives
\begin{equation}\label{eq:fourierconv}
  \fh{k} =  \hh{k} \, \hh{-k} =  \hh{k}^2.
\end{equation}

Taking the origin in the center of the disk of radius $b$ (see the
inset of
\figref{fig:H-of-x}), the available space function $h$ is given by
\begin{equation}\label{eq:defn_of_h}
h(x) = \frac{1}{\modulus{Q}} \left( 1 - 2 \sqrt{b^2 - x^2}
-2 \sqrt{a^2-(\textstyle \frac{1}{2}-x)^2} \right)
\end{equation}
for $x \in [0,1/2)$, and is even and periodic with period $1$.
Here we adopt the convention that $\sqrt{\alpha} = 0$ if
$\alpha<0$ to avoid writing indicator functions explicitly. The
evaluation of the Fourier coefficients of $h$ thus involves
integrals of the form
\begin{equation}\label{eq:bessel}
  \int_0^a \cos zt \, \sqrt{a^2-t^2} \rd t
    = \frac{\pi a}{2z} \, J_1(za),  \qquad (z\neq 0)
\end{equation}
where $J_1$ is the first order Bessel function; this equality
follows from equation (9.1.20) of \cite{AbramowitzS} after a
change of variables.

Hence the Fourier coefficients of $h$ are $\hh{0}=\int_0^1 h(x) =
1$ and, for integer $k\neq 0$,
\begin{equation}\label{eq:fouriercoeffofh}
\ \hh{k} = -\frac{1}{\modulus{Q} . \modulus{k}}
\left[ (-1)^k \, a \, J_1(2 \pi a \modulus{k})
+ b \, J_1(2 \pi b \modulus{k}) \right].
\end{equation}
Note that  $\int_0^1 \phi(x) \rd x = \fh{0} = \hh{0}^2 = 1$, so
that $\phi$ is correctly normalized as a density function on the
torus.

\begin{figure}
\includegraphics{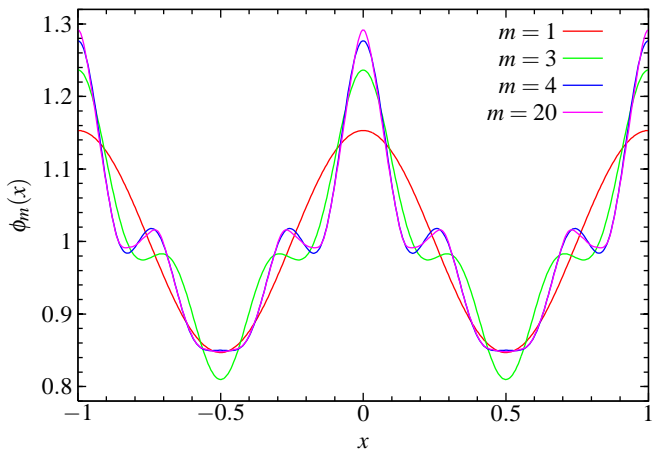}
\caption{\label{fig:partial-sums}(Color online) Partial
sums $\phi_m$ up to $m$ terms of the Fourier series for $\phi$,
with $r=2.3$ and $b=0.5$.}
\end{figure}

In \figref{fig:partial-sums} we plot partial sums $\phi_m$ up to $m$ terms
of the Fourier
series for $\phi$ analogous to \eqref{eq:fourierseries}.
We can determine the degree of smoothness of
$\phi$, and hence presumably of $g_t$, as follows. The
asymptotic expansion of $J_1(z)$ for large real $z$ (equation
(9.2.1) of \cite{AbramowitzS}),
\begin{equation}\label{eq:asympexpansionforJ1}
J_1(z) \sim \sqrt{\frac{2}{\pi z}} \, \cos \left( {
\frac{3\pi}{4}} - z \right) = \bigO{z^{-1/2}},
\end{equation}
shows that $\hh{k} = \bigO{k^{-3/2}}$ and hence
$\fh{k}=\bigO{k^{-3}}$. From the theory of Fourier series (see
e.g.\ \cite[Chap.\ 1]{Katznelson}), we hence have that $\phi$ is
at least $C^1$ (once continuously differentiable).  Thus the
convolution of $h$ with itself is smoother than $h$ is, as intuitively
expected, despite the non-differentiable points of $h$.

We have
checked numerically  the approach of $\int \etator_t(x,y)
\rd y$ to $\phi(x)$, and it appears to be fast, although
the rate is difficult to evaluate,
since a large number of initial conditions are required for
the numerically calculated distribution function to approach
closely the limiting distribution.

\subsection{Structure of displacement distribution}

\begin{figure}
\includegraphics{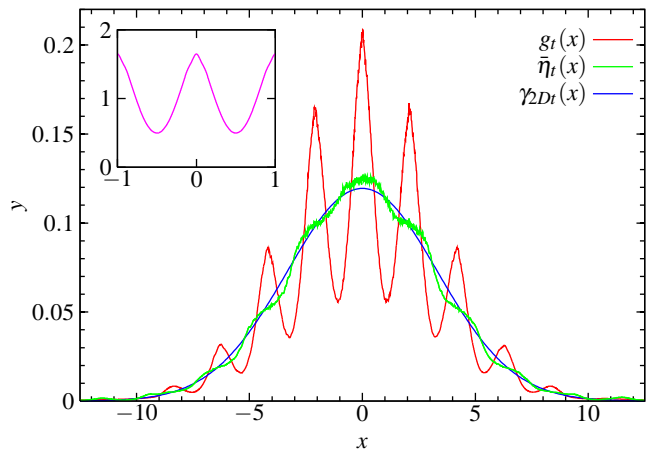}
\caption{\label{fig:demodulate-disp-distn}%
(Color online) Displacement density $g_t$, with demodulated $\etabar_t$
compared to a gaussian
of variance $2D$.  The inset in (a) shows the fine structure function
$\phi$ for these
geometrical parameters.
$r=2.1$; $b=0.2$; $t=50$.
}
\end{figure}

In \figref{fig:demodulate-disp-distn} we plot the
numerically-obtained displacement density $g_t(x)$, the fine
structure function $\phi$ calculated above, and their ratio
$\etabar_t(x)$, for a certain choice of geometrical parameters.
Again the ratio is approximately gaussian, which confirms that the
densities can be regarded as a gaussian shape modulated by the
fine structure $\phi$.

\begin{figure}
\includegraphics{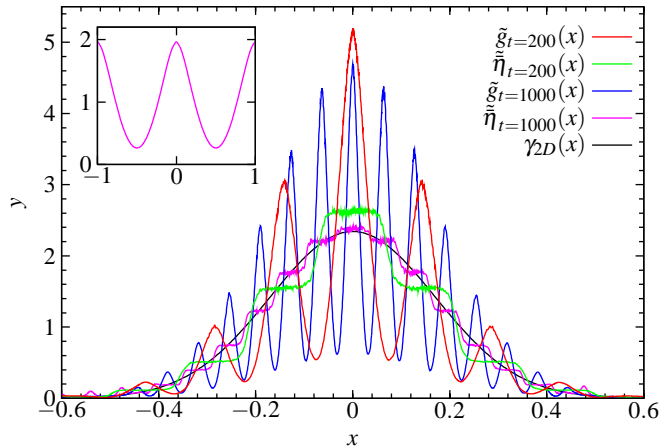}
\caption{\label{fig:demodulate-disp-flat}%
(Color online) Displacement density $g_t$ and demodulated $\etabar_t$, both
rescaled by $\sqrt{t}$, at $t=200$ and $t=1000$, compared to a gaussian
of variance $2D$.
The inset again shows the fine structure function $\phi$.
$r=2.01$; $b=0.1$.
}
\end{figure}

However, if $r$ is close to $2a$, then $\etabar_t$ itself develops a
type of fine structure: it is nearly constant over each unit cell.
 This is shown in \figref{fig:demodulate-disp-flat} for two different times.
We plot both $g_t$ and
 $\etabar_t$, rescaled by $\sqrt{t}$ and compared to a gaussian of
 variance $2D$.  (This scaling is discussed in \secref{sec:clt}.)

 This step-like structure of $\etabar_t$ is related to the validity of
the
\emph{Machta--Zwanzig random walk
approximation}, which gives an estimate of the diffusion
coefficient in regimes where the geometrical structure can be
regarded as a series of traps with small exits
\cite{MZ,KlagesD00, KlagesK02,SandersNext}.
   Having $\etabar_t$ constant across each cell
indicates that  the distribution of particles within the
billiard domain in each cell is uniform, as is needed for the Machta--Zwanzig
approximation to work.

As $r$ increases away from $2a$, the exit size of the traps
increases, and the Machta--Zwanzig argument ceases to give a good
approximation \cite{SandersNext,KlagesD00}. The distribution then
ceases to be uniform in each cell: see
\figref{fig:demodulate-posn-distn}. This may be related to the
crossover to a Boltzmann regime described in \cite{KlagesD00}.

\section{Central limit theorem and rate of convergence}
\label{sec:clt}

We now discuss the central limit theorem as $t \to \infty$ in
terms of the fine structure described in the previous section.

\subsection{Central limit theorem: weak convergence to normal
distribution} \label{subsec:clt-weak-conv}

\begin{figure}
\includegraphics{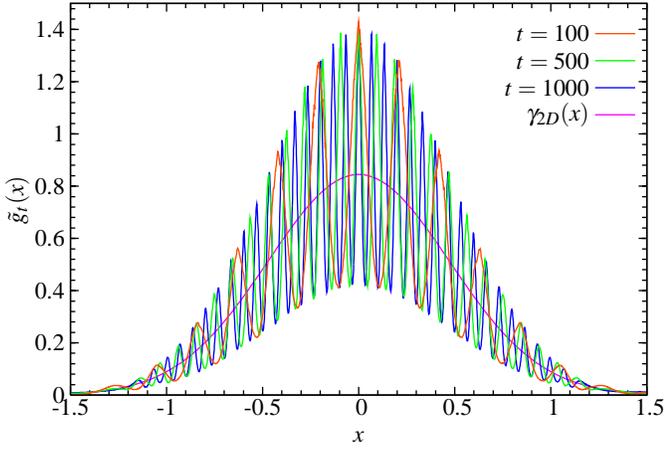}
\caption{\label{fig:rescaled-densities}(Color online) Displacement densities
as in
\figref{fig:time-evol}(b) after  rescaling by $\sqrt{t}$, compared to
a gaussian density with mean $0$ and variance $2D$. $r=2.1$; $b=0.2$.}
\end{figure}

The central limit theorem requires us to consider the densities
rescaled by $\sqrt{t}$, so we define
\begin{equation}\label{}
\tg_t(x) \defeq \sqrt{t} \, g_t(x \sqrt{t}),
\end{equation}
where the first factor of $\sqrt{t}$ normalizes the integral of
$\tg_t$ to $1$, giving a probability density.
\bfigref{fig:rescaled-densities} shows the densities of
\figref{fig:time-evol}(a) rescaled in this way, compared to a gaussian
density with mean $0$ and variance $2D$.
We see that the rescaled densities oscillate within an envelope which
remains approximately
constant, but with an increasing frequency
as $t \to \infty$; they are oscillating around the
limiting gaussian, but do not converge to it pointwise.
See also \figref{fig:demodulate-disp-flat}.

\begin{figure}
\includegraphics{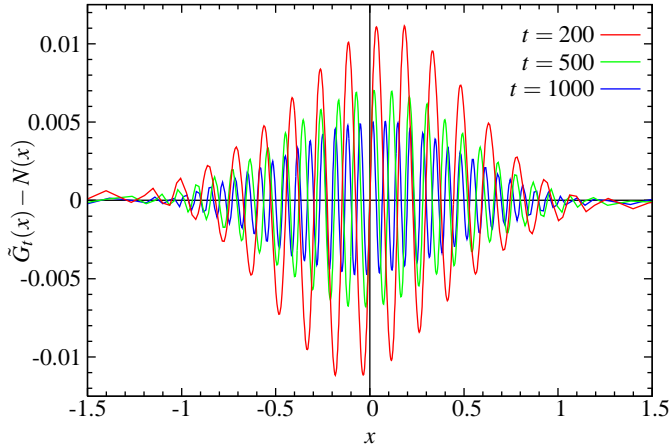}
\caption{\label{fig:rescaled-distn-fns}(Color online) Difference between
rescaled distribution functions and
    limiting normal distribution with variance $2D$. $r=2.1$; $b=0.2$.}
\end{figure}

The increasingly rapid oscillations do, however, cancel out when
we consider the rescaled distribution functions, given by the integral of
the rescaled density functions:
\begin{equation}\label{}
\tG_t(x) \defeq \int_{s=-\infty}^x \tg_t(s) \rd s = G_t(x \,
\sqrt{t}).
\end{equation}
\bfigref{fig:rescaled-distn-fns} shows the difference between the rescaled
distribution
functions and the limiting normal distribution with mean $0$ and
variance $2D$.  We see that
 the rescaled distribution functions do converge
 to the limiting normal, in fact uniformly, as $t \to \infty$; we thus
have \defn{weak} convergence.

Although this is the strongest kind of convergence we can obtain
for the densities $\tilde{g}_t$ with respect to Lebesgue measure,
\figref{fig:demodulate-disp-flat} provides evidence for the
following conjecture: the rescaled densities $\tilde{\etabar}_t$
with respect to the new, modulated measure converge
\emph{uniformly} to a gaussian \emph{density}.  This characterizes
the asymptotic behavior more precisely
than the standard central limit theorem.

\subsection{Rate of convergence} \label{subsec:rate-conv}

Since the $\tG_t$ converge uniformly to the limiting normal
distribution, we can consider
the distance $\supnorm{\tG_t - \normal{2D}}$, where we define the
\defn{uniform norm} by
\begin{equation}\label{}
\supnorm{F} \defeq \sup_{x \in \R} \modulus{F(x)}.
\end{equation}
We denote by $\normal{\sigma^2}$  the
normal distribution function with variance $\sigma^2$, given by
\begin{equation}\label{}
\normal{\sigma^2}(x) \defeq \frac{1}{\sigma \sqrt{2 \pi}}
\int_{s=-\infty}^x \e^{-s^2/2\sigma^2} \rd s,
\end{equation}

\begin{figure}
\includegraphics{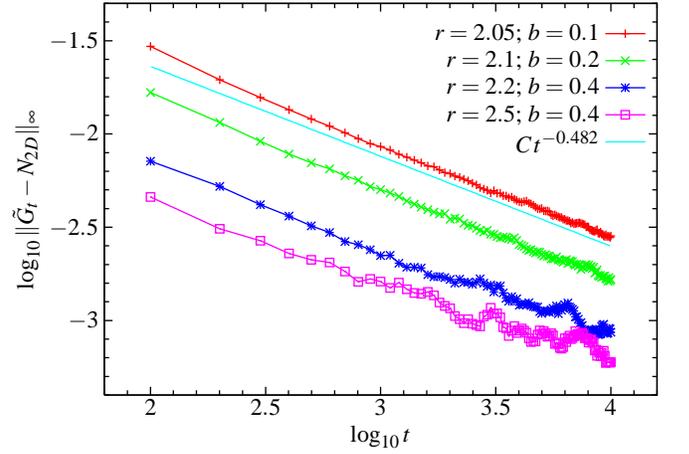}
\caption{(Color online) Distance of rescaled distribution functions
$\tilde{G}_t$
from
limiting normal distribution $N_{2D}$ in log--log plot.  The straight line
is a fit to the large-time decay of the data for $r=2.05$.}
  \label{fig:rate-conv}
\end{figure}

\bfigref{fig:rate-conv} shows a log--log plot of this distance
against time, calculated numerically from the full distribution
functions.  We see that the convergence follows a power law
\begin{equation}\label{}
\supnorm{\tG_t - \normal{2\Dest}} \sim t^{-\alpha},
\end{equation}
with a fit to the data for $r=2.05$ giving a slope $\alpha \simeq 0.482$.
The same decay rate is obtained for
a range of other geometrical parameters, although
the quality of the data deteriorates for larger $r$, reflecting  the fact
that diffusion is faster, so that the distribution spreads further
in the same time.  Since we use the same number
$N=10^7$ of
initial conditions, there is  a lower resolution near $x=0$ where, as shown
in the next section, the maximum
is obtained.

In \cite{Pene} it
was proved rigorously  that $\alpha \ge \frac{1}{6}
\simeq 0.167$ for \emph{any} H\"older continuous observable $f$.
Here we
have considered only the particular H\"older observable
$v$, but for this function we see that the rate of convergence is
much faster than the lower bound proved in \cite{Pene}.

\subsection{Analytical estimate of rate of convergence}
\label{subsec:qual-est-rate-conv}

We now obtain a simple analytical estimate of the rate of
convergence using the  fine structure calculations in
\secref{sec:fine-structure}.

Since the displacement distribution is symmetric, we have
$\tG_t(x=0) = 1/2$ for all $t$. The maximum deviation of $\tG_t$
from $\normal{2\Dest}$ occurs near to $x=0$, where the
density function is furthest from a gaussian, while the fine structure of the
density $\tg_t$ means
 that $\tG_t$ is increasing there (\figref{fig:rescaled-distn-fns}).
 Due to the
oscillatory nature of the fine structure, this maximum thus
occurs at a distance of $1/4$ of the period of oscillation
 from $x=0$.

Since the displacement density has the form $g_t(x) = \phi(x) \etabar_t(x)$,
after rescaling
we have
\begin{equation}\label{}
\tg_t(x) = \phi(x \sqrt{t}) \tilde{\etabar}(x),
\end{equation}
where $\tilde{\etabar}_t(x) \defeq \sqrt{t} \, \etabar_t(x \sqrt{t})$ is the
rescaled slowly-varying part
of $g_t$, and the fine structure at time $t$ is given by
\begin{equation}\label{}
\phi(x \sqrt{t}) = 1 + 2 \sum_{k \in \N} \fh{k} \, \cos(2 \pi k x
\sqrt{t}).
\end{equation}
The maximum deviation occurs at $1/4$ of the period of
$\phi(x\sqrt{t})$, i.e.\ at $x=\frac{1}{4\sqrt{t}}$, so that
\begin{align}\label{}
\supnorm{G_t - N} &\simeq \int_0^{1/4 \sqrt{t}} \sum_{k \in \N}
\fh{k} \,
 \cos(2 \pi k x \sqrt{t})
\rd x \\
&= \frac{1}{\sqrt{t}} \sum_{k \in \N, \, k \text{ odd}} \fh{k}
\frac{(-1)^{(k-1)/2}}{2 \pi k}.
\end{align}
The correction due to the curvature of the underlying gaussian
converges to $0$ as $t \to \infty$, since this gaussian is flat at
$x=0$. Hence $\supnorm{G_t - N} = \bigO{t^{-1/2}}$.

This calculation shows that the fastest possible convergence is a
power law with exponent $\alpha = 1/2$, and provides an intuitive
reason why this is the case. If the rescaled shape function
$\tilde{\etabar}_t$ converges to a gaussian shape at a rate slower
than $t^{-1/2}$, then the overall rate of convergence $\alpha$
could
 be slower than $1/2$.  However, the numerical results in
\secref{subsec:rate-conv} show that the rate is close to $1/2$.
We remark that for an observable
which is not so intimately related to the geometrical structure of
the lattice, the fine structure will in general be more
complicated, and the above argument may no longer hold.

\section{Maxwellian velocity distribution}
\label{sec:maxwell-vel-distn}

In this section we consider the effect of a non-constant
distribution of particle speeds \footnote{The author is indebted
to Hern\'an Larralde for posing this question, and for the
observation that the resulting position distribution may no longer
be gaussian.}.
A Maxwellian (gaussian) velocity distribution was used in polygonal and
Lorentz
channels in \cite{LiHeatLinearMixing} and \cite{AlonsoLorentzChannel},
respectively, in
connection with heat conduction studies. The
  mean square displacement was observed to grow
asymptotically linearly, but  the relationship with the unit speed
situation was not discussed. A more complicated Lorentz gas with a
gaussian distribution was studied in
\cite{KlagesThermostat}.

  We show that the mean square displacement grows asymptotically
linearly in time with the same diffusion coefficient as for the
unit speed case, but that the limiting position distribution may
be \emph{non-gaussian}. For brevity we refer only to the position
distribution throughout
this section; the displacement distribution is similar.

\subsection{Mean square displacement}

Consider a particle located initially at $(\x_0,\v_0)$, where
$\v_0$ has unit speed.  Changing the speed of the particle does
not change the path it follows, but  only  the distance along the
path traveled in a given time.  Denoting by $\flow_v^t(\x_0,\v_0)$
the billiard flow with speed $v$ starting from $\x_0$ and with
initial velocity in the direction of the unit vector $\v_0$, we
have
\begin{equation}
\flow_v^t(\x_0,\v_0) = \flow^{vt}(\x_0, \v_0),
\end{equation}
where the flow on the right hand side is the original unit-speed
flow. If all speeds are equal to $v$, then the radially symmetric
2D position probability density after a long time $t$ is thus
\begin{equation}
\conddensity{p_t}{x,y}{v} = \frac{1}{4\pi Dvt} \exp \left(
\frac{-(x^2+y^2)}{4 D v t} \right),
\end{equation}
giving a radial density
\begin{equation}
\conddensity{p_t}{r}{v} = \frac{r}{2 D v t} \, \exp \left( \frac{-r^2}
{4 D v t} \right).
\end{equation}
(Throughout this calculation we neglect any fine structure.)

If we now have a distribution of velocities with density $p_V(v)$,
then the radial position density at distance $r$ is
\begin{equation} \label{eq:maxwell-posn-density}
\frad_t(r) = \int_{v=0}^\infty \conddensity{p_t}{r}{v} \, p_V(v) \rd v.
\end{equation}
The variance of the position distribution is then given by
\begin{align}
\mean{\x^2} &= \int_{r=0}^\infty r^2 \, \frad_t(r) \rd r \\
&= 4Dt \int_0^\infty v \, p_V(v) \rd v \eqdef 4Dt \bar{v},
\end{align}
where $\bar{v}$ is the mean speed, after interchanging the integrals over
$r$ and $v$.

We thus see that for any speed distribution having a finite mean,
the variance of the position distribution, and hence the
mean square displacement, grows asymptotically linearly with
the same diffusion coefficient as for the uniform speed
distribution, having normalized such that $\bar{v}=1$. We have
verified this numerically with a gaussian velocity distribution:
the mean square displacement is indistinguishable from the unit
speed case even after very short times \cite{MyThesis}.

\subsection{Gaussian velocity distribution}

Henceforth attention is restricted to the case of a gaussian
velocity distribution. For each initial condition, we generate two
independent normally-distributed random variables $v_1$ and $v_2$
with mean $0$ and variance $1$ using the standard Box--Muller
algorithm \cite{NR}, and then multiply by $\sigma$, which is a
standard deviation calculated below. We use $v_1$ and $v_2$ as the
components of the velocity vector $\v$, whose probability density
is hence given by
\begin{equation}\label{}
p(\v)=p(v_1,v_2) = \frac{\e^{-v_1^2/2\sigma^2}}{\sigma \sqrt{2\pi}} \,
\, \frac{\e^{-v_2^2/2\sigma^2}}{\sigma \sqrt{2\pi}} \,
= \frac{\e^{-v^2/2\sigma^2}}{2 \pi \sigma^2} \, ,
\end{equation}
%p(\v)=p(v_1,v_2) &= \frac{1}{\sigma \sqrt{2\pi}} \,
%\e^{-v_1^2/2\sigma^2} \, \frac{1}{\sigma \sqrt{2\pi}} \,
%\e^{-v_2^2/2\sigma^2} \\
%&= \frac{1}{2 \pi \sigma^2} \, \e^{-v^2/2\sigma^2},
where $v \defeq \modulus{\v} = \sqrt{v_1^2+v_2^2}$ is the speed of
the particle.  The speed $v$ thus has density
\begin{equation}\label{}
p_V(v) = \frac{v}{\sigma^2}  \, \e^{-v^2/ 2\sigma^2}
\end{equation}
and mean
$\bar{v} = \sigma \sqrt{\pi/2}$.
 To compare
with the unit speed distribution we require $\bar{v} = 1$, and
hence $\sigma =  \sqrt{2/\pi}$. As before, we distribute the
initial positions uniformly with respect to Lebesgue measure in
the billiard domain $Q$.

\subsection{Shape of limiting distribution}

The position density \eqref{eq:maxwell-posn-density} is a function of time.
However, the gaussian assumption  used to derive that
equation is  valid in the limit when $t \to \infty$, so the central limit
 theorem
rescaling
\begin{equation}
\ftrad_t(r) \defeq \sqrt{t} \, \frad_t(r \sqrt{t})
\end{equation}
eliminates the time dependence in \eqref{eq:maxwell-posn-density},
giving the following shape
for the limiting radial density:
\begin{equation}\label{eq:maxwell-reduced-integral}
\ftrad(r) = \frac{\pi r}{4D} \int_{v=0}^\infty
\exp \left(-\frac{r^2}{4Dv} - \frac{\pi v^2}{4} \right) \rd v \eqdef
\frac{\pi r}{4D} \, I,
\end{equation}
denoting the integral by $I$.
\texttt{Mathematica} \cite{MathematicaBook} can evaluate this integral
explicitly in terms of the \defn{Meijer $G$-function} \cite{Erdelyi}:
\begin{equation}\label{eq:exact-meijer-G}
I = G^{3,0}_{0,3}\left(\frac{\pi r^4}{256 D^2} \right| \left.
\begin{matrix} \text{---} \\ -\textstyle \frac{1}{2},0,0 \end{matrix} \right).
\end{equation}
See \cite{MetzlerKlafter} and references therein for a review of the use
of such special functions in anomalous diffusion.

We can, however, obtain an asymptotic approximation to $I$ from
its definition as an integral, without using any properties of
special functions, as follows.
Define $K(v) \defeq \frac{r^2}{4Dv} + \frac{\pi v^2}{4}$,  the negative
of the argument of the exponential in \eqref{eq:maxwell-reduced-integral}.
Then $K$ has a unique minimum at $\vmin \defeq (r^2/(2 \pi D))^{1/3}$
and we expect the integral to be dominated by the neighborhood of this
minimum.
However, the use of standard asymptotic methods is complicated by the fact
that
as $r \to 0$, $\vmin$ tends to $0$, a boundary
of the integration domain.

To overcome this, we change variables to fix the minimum away from
the domain boundaries, setting $w \defeq v / \vmin$.  Then
\begin{equation}
I = \vmin \int_{w=0}^\infty e^{-\alpha \, L(w)} \rd w,
\end{equation}
where $\alpha \defeq \frac{\pi \vmin^2}{2}$ and $L(w) \defeq \frac{1}{w} +
\frac{w^2}{2}$, with a minimum at $\wmin=1$.
Laplace's method (see e.g.\ \cite{CarrierKrookPearson}) can now be applied,
giving the asymptotic approximation
\begin{equation}\label{eq:asymptotic-expansion-integral}
I \sim \vmin \, e^{-\alpha L(\wmin)} \frac{\sqrt{2\pi}}{\sqrt{\alpha \,
L''(\wmin)}} = \frac{2}{\sqrt{3}} e^{-3\alpha/2},
\end{equation}
valid for large $\alpha$, i.e.\ for large $r$.

Hence
\begin{equation}\label{eq:asymptotic}
\ftrad(r) \stackrel{r \to \infty}{\sim} C \, r \, e^{-\beta \, r^{4/3}},
\end{equation}
where
\begin{equation}\label{eq:defs-of-asymp-consts}
C \defeq \frac{\pi}{2D \sqrt{3}}; \quad
\beta \defeq \frac{3}{2} \left(\frac{\pi}{32 D^2} \right)^{1/3}.
\end{equation}

\subsection{Comparison with numerical results}

\begin{figure}
\includegraphics{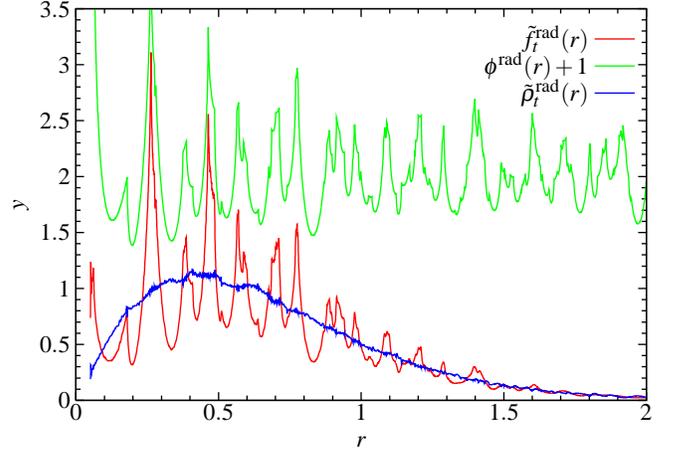}
\caption{\label{fig:maxwell} (Color online) The radial density
function $\ftrad_t$ compared to the numerically calculated  radial
fine structure function $\phirad$, rescaled to converge to $1$ and
then vertically shifted for clarity.  The demodulated radial
density $\rhorad_t$ is also shown. $r=2.3$; $b=0.5$; $t=100$.}
\end{figure}

\bfigref{fig:maxwell} shows the numerical radial position density
$\ftrad_t(r)$ for a particular choice of geometrical parameters.
We wish to demodulate
this as in \secref{sec:fine-structure} to extract the slowly-varying
shape function, which we can then compare to the analytical calculation.

The radial fine structure function $\phirad(r)$ must be calculated
numerically, since no analytical expression is available.  We do
this by distributing $10^5$ points uniformly on a circle of radius
$r$ and calculating the proportion of points not falling inside
any scatterer.  This we normalize so that $\phirad(r)
\to 1$ as $r \to \infty$, using the fact that when $r$ is large,
the density inside the circle of radius $r$ converges to the ratio
$[r^2 - \pi(a^2+b^2)]/r^2$ of available area per unit cell to
total area per unit cell. We can then demodulate $\ftrad_t$ by
$\phirad$, setting
\begin{equation}
\rhorad_t(r) \defeq \frac{\ftrad_t(r)}{\phirad(r \, \sqrt{t})}.
\end{equation}

\bfigref{fig:maxwell} shows the demodulated radial density
$\rhorad_t(r)$ at two times compared to the exact solution
\eqref{eq:maxwell-reduced-integral}--\eqref{eq:exact-meijer-G},
the asymptotic approximation
\eqref{eq:asymptotic}--\eqref{eq:defs-of-asymp-consts}, and the
radial gaussian $\frac{r}{2D} e^{-r^2/2D}$.  The asymptotic
approximation agrees well with the exact solution except at the
peak, while the numerically determined demodulated densities agree
with the exact long-time solution over the whole range of $r$. All
three differ significantly from the gaussian, even in the tails.
We conclude that the radial position distribution is
\defn{non-gaussian}. A similar calculation could be done for the
radial displacement distribution, but a numerical integration
would be required to evaluate the relevant fine structure
function.

An explanation of the non-gaussian shape comes by considering slow particles
which remain close to the origin for a long time, and fast particles which can
travel
further than those with unit speed. The combined effect skews the
resulting distribution in a way which depends on the relative
weights of slow and fast particles.

\begin{figure}
\includegraphics{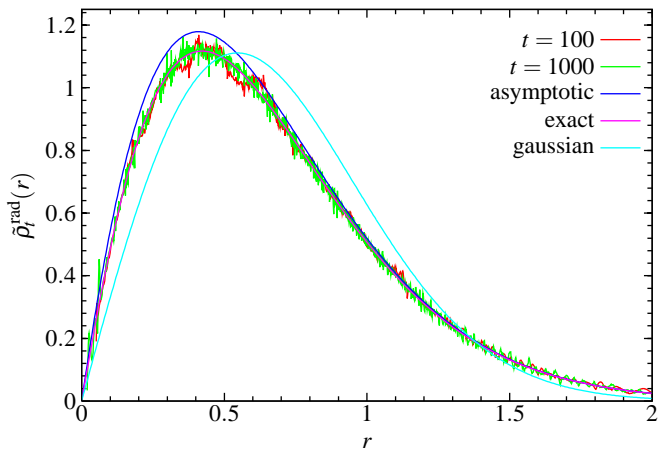}
\caption{\label{fig:maxwell-demod} (Color online) Comparison of
the demodulated radial density $\rhorad_t$ with the exact
Meijer-$G$ representation, the large-$r$ asymptotic approximation,
and the radial gaussian with variance $2D$.}
\end{figure}

\subsection{1D marginal}

\begin{figure}
\includegraphics{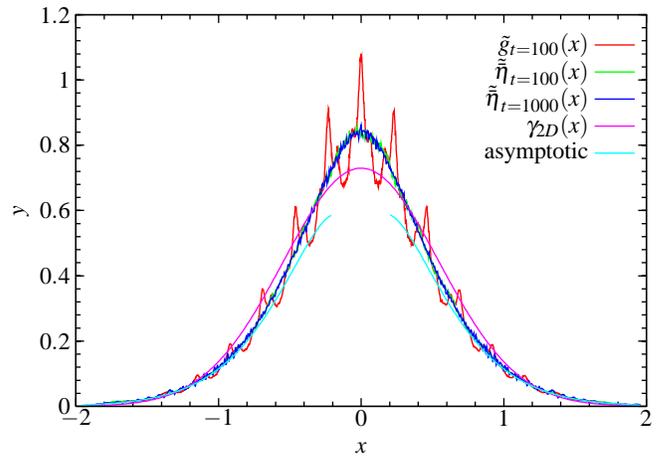}
\caption{\label{fig:maxwell-marginal} (Color online) Rescaled 1D marginal of
the
displacement density $\tilde{g}$ and the demodulated version
$\tilde{\etabar}$ compared to the gaussian with variance $2D$ and
to the asymptotic expression.  The latter is not shown
close to $x=0$, where it drops to $0$. $r=2.3$; $b=0.5$.}
\end{figure}

The 1D marginal in the $x$-direction is shown in
\figref{fig:maxwell-marginal}.  Again there is a significant deviation of the
demodulated density
from a gaussian.
From \eqref{eq:asymptotic}, the 2D density at $(x,y)$ is asymptotically
\begin{equation}\label{eq:2d-asymptotic}
\tilde{f}(x,y) \sim \frac{C}{2\pi} \exp\left[-\beta (x^2+y^2)^{2/3}\right],
\end{equation}
from which the 1D marginal $\tilde{f}(x)$ is obtained by
\begin{equation}\label{eq:asymptotic-marginal}
\tilde{f}(x) = \int_{y=-\infty}^\infty \tilde{f}(x,y) \rd y.
\end{equation}
It does not seem to be possible to perform this integration
explicitly for either the asymptotic expression
\eqref{eq:2d-asymptotic} or the corresponding exact solution in
terms of the Meijer $G$-function.  Instead we perform another
asymptotic approximation starting, from the  asymptotic expression
\eqref{eq:2d-asymptotic}. Changing variables in
\eqref{eq:asymptotic-marginal} to $z
\defeq y/x$ and using the evenness in $y$ gives
\begin{equation}
\tilde{f}(x) \sim \frac{C \modulus{x}}{2\pi} \int_{z=-\infty}^\infty
\exp \left[ -\kappa(1+z^2)^{2/3} \right] \rd z,
\end{equation}
where $\kappa \defeq \beta \modulus{x}^{4/3}$.
 Laplace's method  then gives
\begin{equation}
\tilde{f}(x) \sim \frac{C\sqrt{3}}{\sqrt{8\pi\beta}} \modulus{x}^{1/3}
e^{-\beta \modulus{x}^{4/3}},
\end{equation}
valid for large $x$.
This is also shown in \figref{fig:maxwell-marginal}.  Due to the
$\modulus{x}^{1/3}$ factor, the behavior near $x=0$ is wrong, but
in the tails there is
reasonably good agreement with the numerical results.

\section{Polygonal billiard channel} \label{sec:polyg-bill}

In this section, we apply the previous ideas to a
\emph{polygonal} billiard channel.  Polygonal models differ from
Lorentz gases in that they are not chaotic in the standard sense,
since the Kolmogorov--Sinai entropy and  all Lyapunov exponents
 are zero due
to the weak nature of the scattering from the polygonal sides
\cite{vanBeijerenWeaklyChaoticEntropies}.  Other indicators of the complexity
 of the dynamics
of such systems are required: see e.g.\
\cite{vanBeijerenWeaklyChaoticEntropies} and references therein for
a recent example.

 As far as we aware, there are few rigorous results
on ergodic and statistical properties of these models
\cite{AlonsoPolyg,CasatiProsenMixingTriBilliard}. However, certain
polygonal channels have been found numerically to show
\defn{normal diffusion}, in the sense that our property (a) is
satisfied, i.e.\ the mean squared displacement grows
asymptotically linearly: see e.g.\ \cite{LiHeatLinearMixing,
AlonsoPolyg}. No convincing evidence has so far been given,
however, that property (b), the central limit theorem, can be
satisfied, although it was shown in \cite{DettCoh2} that (c) is
satisfied for some random polygonal billiard models.  Here we show
that polygonal billiards can satisfy the central limit theorem.

\subsection{Polygonal billiard channel model}
\label{subsubsec:polyg-channel-model}

\begin{figure}
\subfigure[]{\label{fig:poly-bill-geom}
\includegraphics{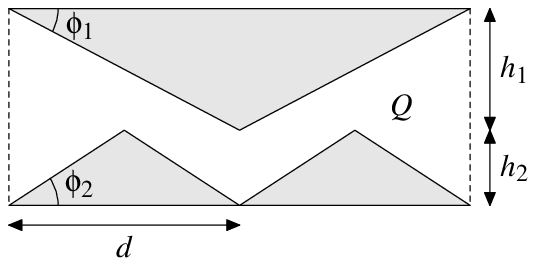}}
\subfigure[]{\label{fig:poly-channel}
\includegraphics{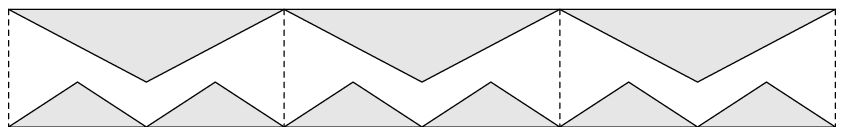}}
  \caption{(a) The geometry of the polygonal billiard unit cell, shown to
scale with $\phi_2=\pi/(2e)$.  (b) Part of the polygonal channel
with the same parameters.}
\end{figure}

We study a polygonal billiard introduced in \cite{AlonsoPolyg}.
The geometry is shown in \figref{fig:poly-bill-geom} and the
channel in \figref{fig:poly-channel}. We fix the angles $\phi_1$
and $\phi_2$ and choose $d$ such that the width of the bottom
triangles is half that of the top triangle.  This determines the
ratio of $h_1$ to $h_2$ in terms of the angles $\phi_1$ and
$\phi_2$.  We then
  require the inward-pointing vertices of each
triangle to lie on the same horizontal line in order to prevent infinite
horizon trajectories, giving $h_1+h_2=h=1$ and $d = h/(\tan
\phi_1 + \textstyle \frac{1}{2} \tan \phi_2)$, with $h_1 = d \tan
\phi_1$ and $h_2 = (d/2) \tan \phi_2$. We remark that in
\cite{AlonsoPolyg} it was  stated that the area $\modulus{Q}=dh$
of the billiard domain is independent of $\phi_2$ when $\phi_1$ is
fixed, but this is not correct, since the expression for $d$ shows
that it depends on $\phi_2$, and we have fixed $h=1$.

In \cite{AlonsoPolyg} the parameters $\phi_1 = \pi(\sqrt{5} -
1)/8$ and $\phi_2 = \pi/q$ were used, with $q \in \N$ and $3 \leq q \leq 9$.
For $q \ge 5$ normal diffusion was found,
whereas for $q=3,4$ it was found that
$\msd_t \sim t^\alpha$ with $\alpha \neq 1$, so that property (a)
is no longer satisfied and we have \defn{anomalous} diffusion.
As far as we are aware, there is as yet no
physical or geometrical explanation for this observed anomalous
behavior, although presumably number-theoretic properties of the
angles are relevant.

We use the same $\phi_1$, but a value of $\phi_2$ which
is irrationally related to $\pi$, namely $\phi_2 = \pi / (2e)
\simeq \pi/5.44$ (where $e$ is the base of natural logarithms),
since there is evidence that mixing properties are stronger for
such irrational polygons \cite{CasatiProsenMixingTriBilliard}.  In
this case we find $\msd_t \sim t^{1.008}$, which we regard as
asymptotically linear, so that
property (a) is again satisfied, with $D = 0.3796 \pm 0.0009$.

\subsection{Fine structure} \label{subsec:poly-fine-structure}

The shape of the displacement density was considered in
\cite{AlonsoPolyg} using histograms, but the results were not
conclusive. Here we use our more refined methods to study the fine
structure of position and displacement distributions and to show
their asymptotic normality.

\begin{figure}
\includegraphics{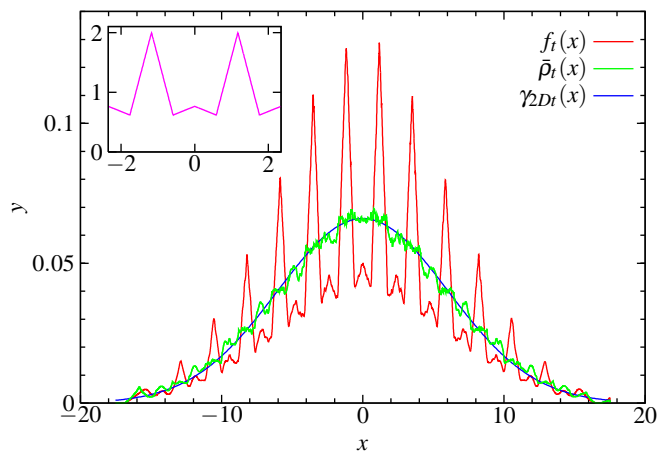}
\caption{\label{fig:poly-posn-density} (Color online) Position
density at $t=50$ in the polygonal model with $\phi_2 = \pi/(2e)$.
The inset shows  $h(x)$ over two periods.}
\end{figure}

\bfigref{fig:poly-posn-density} shows a representative position density
$f_t(x)$.
Following the method of \secref{subsec:unfold-posn-density}, we
calculate the fine structure function $h(x)$ as the normalized height
of available space at position $x$; this is shown in the inset.  We
 demodulate $f_t$ by dividing by $h$ to yield $\rbar_t$, which is again
 close to the gaussian with
variance $2Dt$.

With the same notation as in \secref{subsec:x-disp-density-torus},
 we can also calculate the fine
structure function $\phi$ of the displacement density.
Taking the origin
in the center of the unit cell in \figref{fig:poly-bill-geom}, we
have
\begin{equation}\label{eq:h-for-polyg-bill}
h(x) = \frac{2d}{\modulus{Q}} \left(
x \, \tan \phi_1 + \modulus{x - \textstyle\frac{1}{2} d} \, \tan
\phi_2 \right)
\end{equation}
for $0 \le x \le d$, with $h$ being an even function and having period $2d$.
(The factor of $2d$ in \eqref{eq:h-for-polyg-bill} makes $h$ a density per
unit
length.)
 The Fourier coefficients are $\hh{0} = 1$ and
\begin{equation}\label{}
\hh{k} = \frac{1}{2d} \int_{-d}^d h(x) \, \cos\left( \frac{\pi \, k \, x}{d}
\right) = \frac{1}{\modulus{Q}} \frac{d^2}{\pi^2 k^2} \, l(k)
\end{equation}
for $k \neq 0$, where for $m\in\Z$ we have
\begin{equation}\label{}
l(k) = \begin{cases} 4 \tan(\phi_1), &
\text{if $k$ odd} \\
8 \tan (\phi_2), & \text{if }k = 4m + 2 \\
0, & \text{if }k = 4m.
\end{cases}
\end{equation}

\subsection{Central limit theorem} \label{poly-clt}

\begin{figure}
\includegraphics{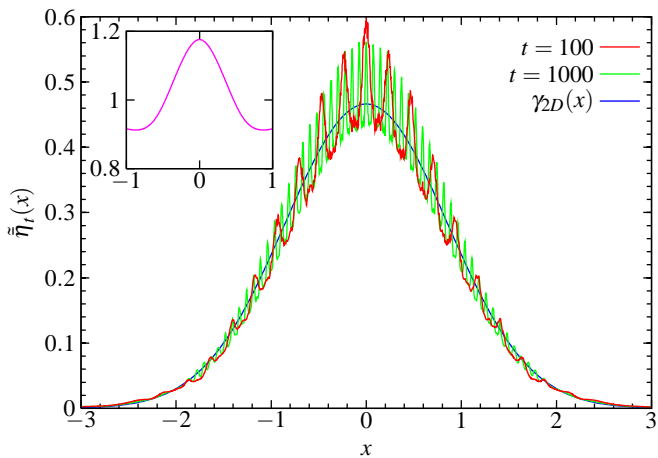}
\caption{\label{fig:phi2-demod}
(Color online) Rescaled displacement denstities compared to the gaussian with
variance $2D$.  The inset shows the function $\phi$ for this geometry.}
\end{figure}

\begin{figure}
\includegraphics{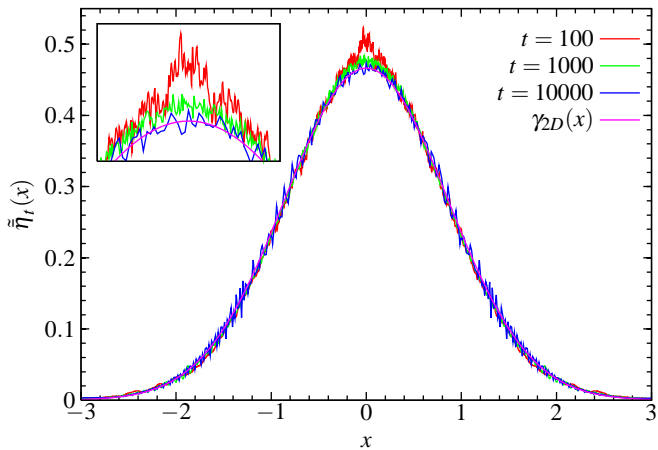}
\caption{\label{fig:poly-demod-rescaled} (Color online) Demodulated densities
$\tilde{\etabar}_t$ for $t=100$, $t=1000$ and $t=10000$,
compared to a gaussian with variance $2D$.  The inset shows a
detailed view of the peak near $x=0$.}
\end{figure}

As for the Lorentz gas, we rescale the densities and distribution functions
by $\sqrt{t}$ to study the convergence to a possible limiting distribution.
Again we find oscillation on a finer and finer scale and
weak convergence to a normal distribution: see \figref{fig:phi2-demod}.
\bfigref{fig:poly-demod-rescaled} shows the time evolution of the
demodulated densities $\tilde{\etabar}_t$.
There is an unexpected peak in the densities near $x=0$ for small times,
indicating some kind of trapping effect; this appears to relax in the long
time limit.  Again we conjecture that we have
uniform convergence of the demodulated densities $\tilde{\etabar}$ to a
gaussian density.

\begin{figure}
\includegraphics{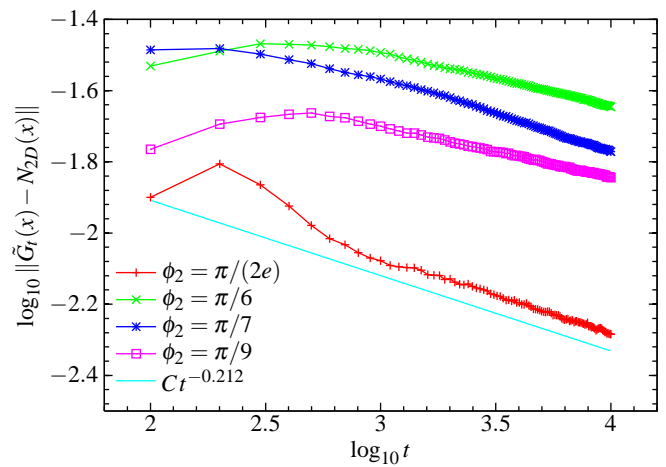}
\caption{\label{fig:poly-clt}(Color online) Distance of the rescaled
distribution
functions from the limiting normal distribution for the polygonal model
with different values of $\phi_2$. The
straight line is a fit to the large-time decay of the irrational case
$\phi_2 = \pi/(2e)$.}
\end{figure}

\bfigref{fig:poly-clt} shows the distance of the rescaled distribution
functions from the limiting
normal distribution,
 analogously to \figref{fig:rate-conv}, for several values of $\phi_2$ for
which the mean square displacement is asymptotically linear. The
straight line fitted to the graph for $\phi_2 = \pi/(2e)$ has
slope $-0.212$, so that the rate of convergence for this polygonal
model is substantially slower than that for the Lorentz gas,
presumably due to the slower rate of mixing in this system. A
similar rate of decay is found for $\phi_2 = \pi/7$, whilst
$\phi_2 = \pi/6$ and $\phi_2=\pi/9$ appear to have a slower decay rate.
Nonetheless, the distance does appear to converge to $0$ for all
these values of $\phi_2$, providing evidence that the
distributions are  asymptotically normal, i.e.\ that the central
limit theorem is  satisfied.

We remark that these convergence rate considerations will be affected
if we have not reached the  asymptotic regime, which would lead to an
incorrect determination of the relevant limiting growth exponent and/or
diffusion coefficient.

\section{Conclusions} \label{sec:conclusions}

We have studied  deterministic diffusion in diffusive billiards in
terms of the central limit theorem. In a 2D periodic Lorentz gas
model, where the central limit theorem is proved, we have shown
that it is possible to understand analytically the fine structure
occurring in the finite-time marginal position and displacement
distribution functions, in terms of the geometry of a unit cell.
Demodulating the observed densities by the fine structure allowed
us to  obtain information about the large-scale shape of the
densities which would remain obscured without this demodulation:
we showed that the demodulated densities are close to gaussian.

We then studied the manner and rate of convergence to the limiting
normal distribution required by the central limit theorem.  We
were able to obtain a simple estimate of the rate of convergence
in terms of the fine structure of the distribution functions.  The
demodulated densities appear to converge uniformly to gaussian
densities,
which is a strengthening of the usual central limit theorem.

We showed that imposing a Maxwellian velocity distribution does
not change the growth of the mean square displacement, but alters
the shape of the limiting position distribution to a non-gaussian
one.

Finally we showed that similar methods can be applied to  a
polygonal billiard channel where few rigorous results are
available, showing that the central limit theorem can be satisfied
by such models, but finding a slower rate of convergence than for
the Lorentz gas.

We believe that our analysis  may
have implications for the escape rate
formalism for calculating transport coefficients (see e.g.\ \cite{GaspBook}),
 where the diffusion equation
with absorbing boundary conditions is used as a phenomenological
model of the escape process from a finite length piece of a
Lorentz gas: analyzing the fine structure in this situation could
provide information about the validity of this use of the
diffusion equation.  We also intend to investigate models
exhibiting anomalous diffusion using the methods presented in this
paper.

\begin{acknowledgments}
I would especially like to thank my PhD supervisor, Robert MacKay,
for his comments, suggestions, and encouragement throughout this
work. I would also like to thank  Leonid Bunimovich, Pierre
Gaspard, Eugene Gutkin, Hern\'an Larralde, Greg Pavliotis, Andrew
Stuart, and Florian Theil for helpful discussions, and
\mbox{EPSRC} for financial support.  The University of Warwick
Centre for Scientific Computing provided computing facilities; I
would like to thank Matt Ismail for assistance with their use.  I
further thank Rainer Klages and an anonymous referee for
interesting comments which improved the exposition of the paper.
\end{acknowledgments}

%\bibliographystyle{apsrev}
%\bibliography{billiards}

\def\cprime{$'$}

\end{document}